\documentstyle[psfig,epsf,aps]{revtex}

\title{Finite-Correlation-Time Effects in the Kinematic Dynamo Problem}
\author{Alexander~A.~Schekochihin\thanks{Present address: 
Blackett Laboratory, Imperial College, 
Prince Consort Rd, London~SW7~2BZ,~U.K.;
electronic mail: sure@pppl.gov} 
and Russell~M.~Kulsrud\thanks{Electronic mail: 
rkulsrud@astro.princeton.edu}\\
{\em Princeton University, P.~O.~Box~451, Princeton, New Jersey 08543}}
\date{19 April 2001}

\def\ssecref#1{Sec.~\ref{#1}}

\def\secref#1{Sec.~\ref{#1}}
\def\apref#1{Appendix~\ref{#1}}
\def\Apref#1{Appendix~\ref{#1}}
\def\exref#1{(\ref{#1})}

\def\eqref#1{Eq.~(\ref{#1})}

\def\figref#1{Fig.~\ref{#1}}

\def\const{{\rm const}}

\def\bea{\begin{eqnarray}}
\def\eea{\end{eqnarray}}

\def\and{{\quad{\rm and}\quad}}

\def\phi{\varphi}

\def\xi{u}
\def\vxi{{\bf\xi}}

\def\({\left(}
\def\){\right)}
\def\[{\left[}
\def\]{\right]}
\def\<{\left\langle}
\def\>{\right\rangle}
\def\l{\left}
\def\r{\right}

\def\d{\partial}
\def\dt{{\d_t}}
\def\diff{{\rm d}}

\def\igamma{\beta}

\def\tkappa{\tilde\kappa}
\def\Idk{{S_d\over(2\pi)^d}\int_0^\infty{\rm d}k}
\def\intk{\int{{\rm d}^d k\over(2\pi)^d}\,}

\def\a{\alpha}
\def\b{\beta}

\def\tZ{{\tilde Z}}
\def\tG{{\tilde G}}
\def\tK{{\tilde K}}
\def\tGamma{{\tilde\Gamma}}

\def\intx{\int{\rm d}^d x}

\def\kbar{{\bar\kappa}}

\def\L{{\hat\Lambda}}
\def\LL{{\hat L}}
\def\MM{{\hat M}}

\def\A{{\hat A}}

\def\vx{{\bf x}}
\def\vy{{\bf y}}
\def\vk{{\bf k}}

\def\vB{{\bf B}}

\def\tcorr{\tau_{\rm c}}
\def\teddy{\tau_{\rm eddy}}

\def\Re{{\rm Re}}

\def\Pr{{\rm Pr}}

\begin{document}

\maketitle

\vskip5mm

\centerline{\tt Published in 
Physics of Plasmas, vol.~8, pp.~4937-4953 (November 2001)} 

\vskip5mm

\begin{abstract}

Most of the theoretical results on the kinematic amplification of 
small-scale magnetic fluctuations by turbulence have been confined 
to the model of white-noise-like ($\delta$-correlated in time) 
advecting turbulent velocity field. 
In this work, 
the statistics of the passive magnetic field in the diffusion-free 
regime are considered for the case when the advecting flow 
is finite-time correlated. 
A new method is developed that allows one to systematically construct 
the correlation-time expansion for statistical characteristics 
of the field such as its probability density function or 
the complete set of its moments. The expansion is valid 
provided the velocity correlation time is smaller than 
the characteristic growth time of the magnetic fluctuations. 
This expansion is carried out up to first order 
in the general case of a $d$-dimensional arbitrarily 
compressible advecting flow. The growth rates for all moments 
of the magnetic-field strength are derived. The effect of 
the first-order corrections due to the finite correlation time 
is to reduce these growth rates. It is shown that 
introducing a finite correlation time leads to the loss of
the small-scale statistical universality, which was present 
in the limit of the $\delta$-correlated velocity field. 
Namely, the shape of the velocity time-correlation profile 
and the large-scale spatial structure of the flow 
become important. The latter is a new effect, that implies, 
in particular, that the approximation of a locally-linear 
shear flow does not fully capture the effect of 
nonvanishing correlation time. 
Physical applications of this theory include the small-scale 
kinematic dynamo in the interstellar medium and protogalactic 
plasmas. 

\end{abstract}

\section{Introduction}

The study of the statistics of magnetic fluctuations excited by a random 
Gaussian white-noise-like advecting velocity field, was pioneered by 
Kazantsev~\cite{Kazantsev}, and, in more recent times, has generated 
a considerable amount of research (see, e.~g., Ref.~\cite{Almighty_Chance} 
and references therein, as well as Refs.~\cite{KA,Gruzinov_Cowley_Sudan,Vergassola,Rogachevskii_Kleeorin,Chertkov_etal_dynamo,BS_metric,AAS_thesis,SBK_review}). 
While much attention has concentrated on  
resistive dynamo problems, most often for very large magnetic Prandtl 
numbers, 
it is well known that the fundamental Zeldovich's, 
or ``stretch--twist--fold,'' 
mechanism of the magnetic-energy amplification (the so-called 
``fast dynamo'') is active regardless of the presence of 
the resistive (diffusive) regularization~\cite{Vainshtein_Zeldovich}. 
If the initial seed magnetic field is concentrated on the scales 
of the same order as the characteristic scales of the advecting 
velocity, the stretching and folding of the magnetic-field 
lines by the random flow
leads to an exponential growth of the magnetic 
fluctuations at scales that decrease exponentially fast, 
until the diffusive scales are 
reached~\cite{Batchelor_vort_analog,KA,AAS_thesis,SBK_review}.  
This scenario is common in astrophysical applications such as 
the turbulence in the interstellar medium or in 
the protogalaxy where the Prandtl number ranges from~$10^{14}$ 
to~$10^{22}$, giving rise to~7 to~11 decades of small (subviscous) 
scales available to the magnetic 
fluctuations~\cite{KA,Kulsrud_etal_proto,SBK_review}. 
In fact, the initial diffusion-free regime may well be the only one 
practically important in such applications as far as the kinematic 
approximation is concerned, since the nonlinear saturation effects 
are likely to set in before the diffusion scales 
are reached~\cite{Kulsrud_lecture}. 
On the fundamental physical level, the diffusion-free regime, 
in which the magnetic-field lines are fully frozen into the flow, 
exhibits most clearly the underlying symmetry properties 
of the passive advection~\cite{BS_metric}. 

With a few notable exceptions (such as Refs.~\cite{DMSR_mean_field,MRS_dynamo_theorem,Gruzinov_Cowley_Sudan,Chertkov_etal_dynamo}), 
the dominant approach in the existing 
literature on the turbulent kinematic dynamo problem 
has been to study the statistics of passive magnetic fields advected 
by a flow $\vxi(t,\vx)$ whose two-time correlation 
function is approximated by a $\delta$ function, 
$\<\vxi(t)\vxi(t')\>\propto\delta(t-t')$. This white-noise property of 
the velocity greatly simplifies matters: the evolution equations 
for such statistical quantities as the correlation functions and 
probability density function of the magnetic field can be derived 
in closed form and yield themselves to exact solution.  

In this paper, we relax the white-noise assumption and explore 
the effects that arise when a finite-time correlated velocity 
field is introduced. This immediately raises the level 
of difficulty associated with solving the statistical 
problem. Within the theoretical framework adopted here, 
the difficulty can be described in the following terms. 
In the zero-correlation-time approximation, 
one essentially has to deal with only one closed differential equation that 
fully determines the desired statistics. Allowing for a finite 
velocity correlation time leads to an infinite number of interlinked
integro-differential equations involving time-history integrals. 
These equations form an infinite open hierarchy that formally 
constitutes the exact description of 
the problem (these matters are explained in more detail 
in~\ssecref{KT_method_hierarchy}).
Solving this hierarchy in its entirety without additional assumptions 
appears to be an impossible task. 
The most obvious way to make progress is clearly to try 
a perturbative approach, i.e., to consider the kinematic dynamo  
problem with an advecting field 
whose correlation time is short, but finite. 
If the correlation time~$\tcorr$ is assumed to be small, one can 
expect to be able 
to construct an expansion in the powers of~$\tcorr$ (in what follows, 
we will frequently refer to it as {\em the $\tau$~expansion})
and calculate corrections to the growth rates of the moments 
of the magnetic field. This is the program that we undertake here. 

We consider the one-point statistics of the passive magnetic field 
in the diffusion-free regime. 
In this context, the infinite hierarchy we have mentioned 
above interrelates the one-point probability density function~(PDF)
of the magnetic field and an infinite set of response functionals.  
These are averaged multiple functional derivatives of 
the magnetic field with respect to the velocity field and its 
gradients. 
We develop a {\em functional expansion method} that allows us 
to calculate successive terms in the $\tau$~expansion 
and derive in a closed form 
a Fokker--Planck equation for the one-point~PDF of the magnetic field. 
We limit ourselves to advancing 
the expansion one order beyond the zero-correlation-time 
approximation. The result is a set of corrections to the growth 
rates of all moments of the magnetic field. These corrections 
are negative, so the growth rates are reduced. 
 
The expansion is carried out assuming 
that the velocity correlation time is small and keeping 
the time integral of the velocity correlation 
function fixed. The latter constraint ensures that the dynamo 
growth rate remains finite when the correlation time vanishes. 
An alternative way, which is sometimes deemed preferable on physical 
grounds (see, e.g, Ref.~\cite{Kinney_etal_2D}), 
is to fix the total energy of the velocity field. 
Since a $\delta$-correlated velocity field must necessarily 
possess infinite energy, fixing the energy at a finite value 
leads to vanishing of the growth rates when~$\tcorr=0$.
The {\em relative} ordering of the terms in 
the expansion is, however, the same, regardless of what 
is kept fixed, so the technical side of the expansion method 
is unaffected. 

Our expansion technique 
will be given detailed treatment in the body of this paper. 
Here, let us rather discuss the finite-correlation-time effects 
that can be distilled on the basis of our approach.
As it turns out, a number of new interesting phenomena 
manifest themselves already at the level of 
the short-but-finite-correlation-time approximation. 

In the case of the $\delta$-correlated advecting flow, 
the one-point statistics of the passive magnetic field are 
{\em universal} in the sense that they only depend on 
one small-scale property of the velocity: 
the time integral of the one-point correlation tensor of its gradients, 
$\int\diff t\<\nabla\vxi(t)\nabla\vxi(0)\>$. 
The essential novelty in the case of finite correlation time 
is that this small-scale universality is lost on two accounts.  

First, the $\tau$~expansion exhibits a sensitive dependence 
on the specific shape of the time-correlation profile of the 
velocity field (in recent literature, this was first explicitly 
pointed out by Boldyrev~\cite{Boldyrev_tcorr}; 
see also Refs.~\cite{vanKampen,vanKampen_review}). 
Namely, multiple time integrals of products 
of velocity correlation functions enter the expressions for 
the expansion coefficients. Choosing different correlation 
profiles leads to order-one changes in the values of these 
coefficients. The root of this nonuniversality lies in the topology 
of the vertex-correction diagrams that contribute to the 
orders higher than the zeroth in 
the $\tau$~expansion (see~\ssecref{tau_discussion}).

Second, the first-order terms of the $\tau$~expansion 
feature a part that arises from the {\em fourth-order} 
derivatives of the velocity correlation function, i.e., from 
the second derivatives of the velocity field. In the one-point statistical 
approach, this is the first manifestation 
of the more general tendency that introducing finite correlation times 
brings into play the large-scale structure of the velocity field. 
A related effect is the loss of Galilean invariance due to 
the fact that the expansion terms also depend on the actual {\em energy} 
of the velocity field, i.e., on the rms value of the sweeping velocity.
Indeed, now that the trajectories of the fluid elements have a ``memory'' 
of themselves, which extends approximately one~$\tcorr$ back in time, 
we should naturally expect that there will appear 
an effective ``correlation length'' 
of the velocity (in what regards the one-point statistics of the fields 
it advects) approximately equal to~$\xi\tcorr$. Therefore, the one-point 
statistics of the passive fields now depend not only on 
the instantaneous velocity difference between two fluid particles 
that meet at a given time (i.e., the velocity gradient at a point), 
but also on the velocity that swept them into place and on 
the variation of the velocity gradient over the correlation length.  
This appearance of first-order corrections due to the second derivatives 
of the flow is a new effect, which indicates, in particular, 
that the customary approximation used in the Batchelor 
regime, where the advecting velocity is assumed to be locally 
linear~\cite{Batchelor,ZRMS_linear}, is only justified for 
the $\delta$-correlated-in-time advecting fields. 

Such are the main qualitative consequences of introducing 
a finite-time-correlated velocity field into the kinematic dynamo 
problem (or, in general, any passive-advection model). 
A few words are in order as to the quantitative 
impact of a finite correlation time on the dynamo action. 
As we have already mentioned, the effect of the first-order 
corrections is to reduce the growth rates of all moments of 
the magnetic field. Besides the nonuniversal dependence 
on the spatial and temporal structure of the velocity 
correlation function, the reduction depends in a universally 
calculable way on the usual set of parameters: 
the order of the moment, the dimension of space, 
and the degree of compressibility of the flow. 
The overall magnitude of this reductive effect is measured 
by the expansion parameter, which is of the order 
of~$\tcorr\gamma$, where $\gamma$~is the growth rate of 
the magnetic energy. It is not hard 
to demonstrate (see~\ssecref{growth_rates_expanded}) 
that~$\tcorr\gamma d\sim(\tcorr/\teddy)^2$, where 
$\teddy$~is the ``eddy-turnover'' time of the advecting 
turbulent velocity field and $d$~is the dimension of space. 
In a standard Kolmogorov-type turbulence setting, one would, of course, 
expect any such approximation to be valid at best marginally, since 
$\tcorr \sim \teddy$. Astrophysical plasmas offer more 
variety in this respect, as their driving forces (typically supernova 
explosions) can, in fact, decorrelate faster than the turbulent eddies 
turn over~\cite{Almighty_Chance}. In any event, 
the small-$\tcorr$ expansion does not offer much 
more than qualitative, or, at best, semiquantitative, 
information about the way the dynamo action is modified 
by the finiteness of the correlation time. It is, of course, 
clear that introducing a finite correlation time cannot 
altogether suppress the fast-dynamo 
mechanism~\cite{MRS_dynamo_theorem,Chertkov_etal_dynamo}. 
On the other hand, 
our conclusion that some reduction of the growth rate should be 
expected, is corroborated by numerical 
evidence~\cite{Chandran_tcorr,Kinney_etal_2D,Chou} 
that suggests a reduction of about~40\% to~50\%. 
In fact, in~\secref{sec_one_eddy}, we offer a semiquantitative 
evaluation of the finite-$\tcorr$ correction to the growth rate 
which yields a reduction of approximately~40\% in the three-dimensional 
case and for~$\tcorr\sim\teddy$. Of course, this is at best 
just an indication of the well-behaved character of our 
expansion, rather than a truly solid quantitative confirmation of~it. 

The literature on the $\tau$~expansion and finite-correlation-time 
effects is not extensive. 
Kliatskin and Tatarskii~\cite{Kliatskin_Tatarskii} were the first 
to propose the hierarchy of equations for the response functionals 
as a starting point for a method of successive approximations 
as applied to the description of waves propagating in a medium 
with random inhomogeneities. Vainshtein~\cite{Vainshtein_KT} 
applied this method to the mean-field 
kinematic dynamo theory. The Kliatskin--Tatarskii method 
and its relation to our functional expansion method are discussed 
at the end of~\ssecref{KT_method_exp}. 
Van~Kampen~\cite{vanKampen} and Terwiel~\cite{Terwiel} developed 
the so-called cumulant expansion method; van~Kampen's review 
article~\cite{vanKampen_review} 
also contains a good critical survey of other $\tau$-expansion 
schemes predating his work. 
His method was later 
applied in the kinematic-dynamo context by Knobloch~\cite{Knobloch} 
and Chandran~\cite{Chandran_tcorr}. Their treatment was Lagrangian and 
did not include any effects due to the explicit spatial dependence  
in the induction equation. Consequently, the nonuniversality of 
the $\tau$~expansion with respect to the spatial structure of 
the velocity correlator was not captured. 
The van~Kampen method is discussed in detail in~\ssecref{VK_method}.
Parallel to our development of the functional expansion method, 
Boldyrev~\cite{Boldyrev_tcorr} proposed a $\tau$-expansion 
method that was based on the exact 
solution of the induction equation in the Lagrangian frame 
and offered a way to calculate the second moment of the magnetic 
field that elicited the nonuniversal character of 
the $\tau$~expansion with respect to both temporal and spatial 
properties of the velocity correlation tensor.   
Molchanov, Ruzmaikin, and Sokoloff~\cite{MRS_dynamo_theorem} 
considered the statistics of the kinematic dynamo in 
a renovating flow using the formalism of infinite products 
of random matrices. 
(See also~Ref.~\cite{DMSR_mean_field} for the treatment 
of the kinematic mean-field dynamo in a renovating flow.) 
A version of their approach was later advanced 
by Gruzinov, Cowley, and Sudan~\cite{Gruzinov_Cowley_Sudan}. 
Considerable progress was achieved in a nonperturbative way by 
Chertkov {\em et al.}~\cite{Chertkov_etal_scalar2D}, 
who studied the passive-scalar problem in two dimensions for arbitrary 
velocity correlation times. However, their method only works in 
the two-dimensional case. 
 
Thus, while we now seem to have a fairly good understanding of 
the structure of the $\tau$~expansion and such qualitative 
features as the loss of the small-scale universality, 
an adequate nonperturbative theory of the kinematic dynamo 
and passive advection in finite-time-correlated turbulent velocity 
fields remains an open problem.  

This paper is organized in the following way. 
In~\secref{sec_tau_exp}, our functional expansion method 
is systematically developed on the example of the simplest 
available passive-advection problem: that of the Lagrangian 
passive vector in an incompressible flow. In this model, 
no explicit spatial dependence is present. 
In~\ssecref{KT_method_hierarchy}, \ssecref{KT_method_Mark}, 
and~\ssecref{KT_method_exp}, 
we present a functional formalism that allows one to systematically 
construct successive terms in the $\tau$-expanded 
Fokker--Planck equation. 
The dependence of the expansion coefficients on the 
specific functional form of the velocity time correlation
profile emerges.
The expansion is carried out up to the first order in~$\tcorr$.
In~\ssecref{VK_method}, our method is compared with 
the van~Kampen cumulant expansion method~\cite{vanKampen}. 
We ascertain that results obtained via the van~Kampen method 
are consistent with ours. 
Finally, in~\ssecref{tau_discussion}, we discuss the underlying 
structure of the $\tau$~expansion in diagrammatic terms. 
In~\secref{sec_KT_dynamo}, the general arbitrarily compressible 
space-dependent dynamo problem is solved with the aid 
of the functional expansion. At this level, the nonuniversality 
with respect to the spatial structure of the velocity correlations, 
as well as the loss of Galilean invariance, become evident. 
In~\ssecref{hierarchy}, we explain 
the emergence of an infinite hierarchy of equations for the characteristic 
function and various averaged response functionals of the magnetic field 
in the passive dynamo problem with finite-time-correlated advecting flow. 
The hierarchy is advanced up to the emergence of the second-order 
response functions.
In~\ssecref{tau_expansion}, we construct the $\tau$~expansion 
up to first order in the correlation time, which leads to a closed equation 
for the characteristic function of the magnetic field.  
In~\ssecref{Fokker_Planck_tcorr}, we derive the Fokker--Planck 
equation for the one-point~PDF of the magnetic-field strength 
valid to first order in the correlation time. The distribution 
is lognormal. 
In~\ssecref{growth_rates_expanded}, we calculate the rates of 
growth of all moments of the magnetic field with (negative) first-order 
corrections. 
Finally, in~\secref{sec_one_eddy}, 
we give a semiquantitative argument 
that relates the expansion parameter to the ratio of 
the correlation and eddy-turnover times of the velocity field.  
We also evaluate the finite-$\tcorr$ reduction of the magnetic-energy 
growth rate in a model incompressible turbulence consisting of 
eddies of a fixed size. 
In~\Apref{ap_k_space}, we provide the basic relations that 
allow one to express the results we have obtained in the configuration 
space in terms of the spectral characteristics of the velocity field. 
Some of the more cumbersome technical details of the $\tau$~expansion 
are exiled to~\Apref{ap_response2}.

\section{The Gaussian Functional Expansion Formalism}
\label{sec_tau_exp}

In this Section, we explain the Gaussian functional method 
for constructing the short-correlation-time expansion for 
passive~advection problems. 
Working out such expansions for specific problems 
often involves a fair amount of algebra, 
which tends to obscure the otherwise transparent 
ideas behind them. 
In an attempt at the maximum possible clarity of exposition, 
we first consider a model that, 
while preserving most of the essential features of the 
passive-advection problems, offers much greater technical simplicity. 
Namely, let us consider the following stochastic equation 
in $d$~dimensions:
\bea
\label{model_B}
\dt B^i = \sigma^i_k B^k.
\eea
All the fields involved explicitly depend on time only. 
The specific initial distribution of~$B^i$ is not important 
for the derivation or the validity of the results below. 
Spatial isotropy is always assumed. 
The matrix field~$\sigma^i_k(t)$ is Gaussian with zero mean 
and a given two-point correlation tensor:
\bea
\label{model_sigma}
\bigl<\sigma^i_k(t)\sigma^j_l(t')\bigr> = T^{ij}_{kl}\kappa(t-t'),
\qquad\qquad\\
\nonumber
T^{ij}_{kl} = \delta^{ij}\delta_{kl} + a\bigl(\delta^i_k\delta^j_l 
+ \delta^i_l\delta^j_k\bigr),\qquad a = -1/(d+1).
\eea
These equations can be interpreted to describe the evolution of 
a passive magnetic field in a Lagrangian frame, where 
the Lagrangian advecting velocity field is Gaussian 
and incompressible, and the tensor~$\sigma^i_k$ is its gradient matrix. 
In the more general context of the theory of passive advection, 
equations~\exref{model_B} and~\exref{model_sigma} model the stochastic 
dynamics of a vector connecting two Lagrangian tracer particles in 
an ideal fluid.

We assume that the temporal correlation function~$\kappa(t-t')$ 
of~$\sigma^i_k(t)$ has a certain characteristic width~$\tcorr$, 
i.e., the field~$\sigma^i_k(t)$ possesses a correlation time~$\tcorr$.
Our task in this section is to construct an expansion 
of the statistics of~$B^i(t)$ in powers of~$\tcorr$, which is 
assumed to be small. The limit~$\tcorr\to0$ 
ought to be taken in such a way that the time integral of the 
correlation function is kept constant:
\bea
\label{kbar_const}
\int_0^\infty\diff\tau\,\kappa(\tau) = {\kbar\over2} = {\rm const}
\quad{\rm and}\quad\tcorr\kbar \ll 1.
\eea
The white-noise limit of zero correlation time is realized 
by setting~$\kappa(\tau)=\kbar\delta(\tau)$. 

The first step in our averaging scheme is to define 
the characteristic function of the field~$B^i(t)$,
\bea
\label{model_Z}
Z(t;\mu) = \bigl<\tZ(t;\mu)\bigr> 
= \bigl<\exp\bigl[i\mu_i B^i(t)\bigr]\bigr>.
\eea
Here and in what follows, the overtildes designate unaveraged 
random functions. Upon differentiating~$\tZ(t;\mu)$ with respect 
to time and making use of~\eqref{model_B}, we obtain a new stochastic 
equation:
\bea
\label{model_eq_Ztilde}
\dt\tZ = \sigma^i_k\mu_i{\d\over\d\mu_k}\,\tZ 
= \L^k_i\sigma^i_k\tZ,
\eea 
where the auxiliary operator~$\L^k_i$ has been introduced 
for the sake of notational compactness. 

Our objective now is to learn how to obtain a closed equation 
for the averaged characteristic function~$Z(t;\mu)$, i.e., 
how to average~\eqref{model_eq_Ztilde} when~$\sigma^i_k$ has 
a nonzero correlation time. The inverse Fourier transform 
(with respect to~$\mu_i$) of the resulting equation 
will be the Fokker--Planck equation for the PDF~$P(t;\vB)$ of 
the passive field~$B^i$.

\subsection{The Hierarchy of Response Functions}
\label{KT_method_hierarchy}

We start the construction of the functional expansion 
by developing an exact formalism that describes 
the one-point statistics of the field~$B^i(t)$. 
Let us average both sides of~\eqref{model_eq_Ztilde} 
and ``split'' the mixed average that arises on the right-hand 
side with the aid of the well-known 
Furutsu--Novikov (or ``Gaussian-integration'') 
formula~\cite{Furutsu,Novikov}:
\bea
\label{model_eq_Z}
\dt Z(t) = \L^k_i\bigl<\sigma^i_k(t)\tZ(t)\bigr> 
= \L^k_i T^{i\b}_{k\a}\int_0^t\diff t'\,\kappa(t-t')G^\a_\b(t|t'),
\eea
where we have suppressed the~$\mu$'s in the arguments, 
used the formula~\exref{model_sigma} for the second-order 
correlation tensor of~$\sigma^i_k(t)$, and introduced 
the {\em averaged first-order response function:} 
\bea
\label{def_G1}
G^\a_\b(t|t') = \bigl<\tG^\a_\b(t|t')\bigr> = 
\biggl<{\delta\tZ(t)\over\delta\sigma^\b_\a(t')}\biggr>.
\eea
This function is subject to the causality constraint: 
$G^\a_\b(t|t')=0$ if~$t'>t$ [hence the upper integration limit 
in~\eqref{model_eq_Z}]. 
Integrating~\eqref{model_eq_Ztilde} from~$0$ to~$t$, taking 
the functional derivative~$\delta/\delta\sigma^\b_\a(t')$ of 
both sides, averaging, setting~$t'=t$, and taking causality into 
account, we~get
\bea
\label{G1_tt}
G^\a_\b(t|t) = \L^\a_\b Z(t).
\eea
We have thus obtained the equal-time form of~$G^\a_\b(t|t')$.
In order to determine the response function at~$t'<t$, 
we simply take the functional derivative~$\delta/\delta\sigma^\b_\a(t')$ 
of both sides of~\eqref{model_eq_Ztilde} and find that 
each element of the unaveraged tensor~$\tG^\a_\b(t|t')$ 
satisfies an equation identical in form to~\eqref{model_eq_Ztilde}. 
Upon averaging this, we obtain an evolution equation 
for~$G^\a_\b(t|t')$ subject to the initial condition~\exref{G1_tt} 
at~$t=t'$:
\bea
\label{G1_eq}
\dt G^\a_\b(t|t') = \L^l_j\bigl<\sigma^j_l(t)\tG^\a_\b(t|t')\bigr>.
\eea
This equation can now be handled in the same fashion 
as~\eqref{model_eq_Ztilde}, the average on the right-hand side 
split via the Furutsu--Novikov formula in terms of the correlation 
tensor of~$\sigma^j_l(t)$ and the appropriately defined 
second-order averaged response 
function~$G^{\a_1\a_2}_{\b_1\b_2}(t|t_1,t_2)$. 
At equal times, the latter can be expressed in terms of~$G^\a_\b(t|t')$ 
just as~$G^\a_\b(t|t)$ was expressed in terms of~$Z(t)$. 
At different times, we obtain the evolution equation 
for~$G^{\a_1\a_2}_{\b_1\b_2}(t|t_1,t_2)$ by taking the functional 
derivative of the equation for~$\tG^\a_\b(t|t')$ and averaging. 

An infinite linked hierarchy can be constructed by further 
iterating this procedure and introducing response functions 
of ascending orders. Let us give the general form of this 
hierarchy. Define the {\em $n$th-order averaged response function:} 
\bea
\label{def_Gn}
G^{\a_1\dots\a_n}_{\b_1\dots\b_n}(t|t_1,\dots,t_n) = 
\biggl<{\delta\tZ(t)\over\delta\sigma^{\b_1}_{\a_1}(t_1)\dots
\delta\sigma^{\b_n}_{\a_n}(t_n)}\biggr>.
\eea
This function has two essential properties: 
(i) it is causal: $G^{\a_1\dots\a_n}_{\b_1\dots\b_n}(t|t_1\dots t_n)=0$
if any~$t_i>t$; (ii) it remains invariant under all 
simultaneous permutations of the times $t_1,\dots,t_n$ 
and indices $\a_1,\dots,\a_n$, $\b_1,\dots,\b_n$, 
which correspond to changes of the order of functional 
differentiation in the definition~\exref{def_Gn}.
The $n$th-order response function satisfies the following 
recursive relations: if~$t_1,\dots,t_n<t$,
\bea
\label{Gn_eq}
\dt G^{\a_1\dots\a_n}_{\b_1\dots\b_n}(t|t_1\dots t_n) = 
\L^k_i T^{i\b_{n+1}}_{k\a_{n+1}}\int_0^t\diff t_{n+1}\,\kappa(t-t_{n+1})
G^{\a_1\dots\a_{n+1}}_{\b_1\dots\b_{n+1}}(t|t_1\dots t_{n+1});
\eea
if, say, $t_n=t$ and $t_1,\dots,t_{n-1}\le t_n$,   
\bea
\label{Gn_tt}
G^{\a_1\dots\a_{n-1}\a_n}_{\b_1\dots\b_{n-1}\b_n}(t|t_1\dots t_{n-1},t) = 
\L^{\a_n}_{\b_n} 
G^{\a_1\dots\a_{n-1}}_{\b_1\dots\b_{n-1}}(t|t_1\dots t_{n-1}).
\eea
The hierarchy is ``forward'' at different times 
and ``backward'' at equal times. 
The characteristic function~$Z(t)$ is formally treated 
as the zeroth-order response function.

\subsection{The White-Noise Approximation}
\label{KT_method_Mark}

The white-noise approximation is obtained by 
setting~$\kappa(t-t')=\kbar\delta(t-t')$. We are then left 
with just~\eqref{model_eq_Z}, where the time history 
integral reduces to~${1\over2}\kbar G^\a_\b(t|t)$, which is substituted 
from~\eqref{G1_tt}. This produces a closed evolution 
equation for~$Z(t)$. The Fourier transform of it is
the Fokker--Planck equation for the~PDF of~$B^i$ at time~$t$
in the $\delta$-correlated regime:
\bea
\label{FPEq_delta}
\dt P(t) = {\kbar\over2}\,T^{i\b}_{k\a} \L^k_i \L^\a_\b P(t) 
= {\kbar\over2}\,\LL P(t).
\eea
In order to be not overly burdened by notation, 
we typically use the same symbol for denoting an operator 
in the Fourier space of the~$\mu$'s and its analog in the configuration 
space of the~$B$'s. This should lead to no confusion, as 
the context will always be clear. Thus, 
\bea
\L^k_i = \mu_i{\d\over\d\mu_k} = - {\d\over\d B^i} B^k 
= - \(\delta^k_i + B^k{\d\over\d B^i}\). 
\eea
Due to isotropy, the probability density function~$P$ depends 
on the absolute value~$B=|\vB|$ only.
The operator~$\LL$ in~\eqref{FPEq_delta} can therefore 
be written in the following isotropic form:
\bea
\label{def_L_iso}
\LL = T^{i\b}_{k\a} \L^k_i \L^\a_\b = 
{d-1\over d+1}\[B^2{\d^2\over\d B^2} + (d+1)B{\d\over\d B}\],
\eea 
which turns~\eqref{FPEq_delta} into the familiar 
Fokker--Planck equation for the one-point~PDF of the 
magnetic field in the kinematic $\delta$-correlated 
dynamo problem taken for the incompressible 
flow~\cite{BS_metric,AAS_thesis}. 
The resulting distribution is lognormal and the moments of~$B$~satisfy 
\bea
\label{expected_outcome}
\dt\<B^n\> = 
{1\over2}\,{d-1\over d+1}\,n\(n+d\)\kbar\<B^n\>.
\eea
This is the expected outcome, because, 
as was shown in Ref.~\cite{BS_metric},
the Lagrangian and Eulerian statistics are the same 
for the $\delta$-correlated incompressible flow. 

Thus, the solution in the $\delta$-correlated limit is quite 
elementary. Things become much more complicated once the 
white-noise assumption is relaxed and a nonzero, however small, 
velocity correlation time is introduced.

\subsection{The Recursive Expan\-sion}
\label{KT_method_exp}

In order to construct an expansion in small correlation time, 
it is convenient to combine the equations~\exref{Gn_eq} 
and~\exref{Gn_tt} into one recursive integral relation 
that expresses the $n$th-order response function in terms 
of its immediate precursor and its immediate successor:
\bea
\nonumber
G^{\a_1\dots\a_n}_{\b_1\dots\b_n}(t|t_1\dots t_n) =
\L^{\a_n}_{\b_n} 
G^{\a_1\dots\a_{n-1}}_{\b_1\dots\b_{n-1}}(t_n|t_1\dots t_{n-1})
\qquad\qquad\qquad\qquad\\ 
\label{Gn_rec_rln}
+\,\,\L^k_i T^{i\b_{n+1}}_{k\a_{n+1}}
\int_{t_n}^t\diff t'\int_0^{t'}\diff t_{n+1}\,\kappa(t'-t_{n+1})
G^{\a_1\dots\a_{n+1}}_{\b_1\dots\b_{n+1}}(t'|t_1\dots t_{n+1}).
\eea 
The above relation is exact and 
valid for $t_1,\dots,t_{n-1} \le t_n \le t$.
Due to the permutation symmetry of the response functions, 
this does not limit the generality.   
The desired expansion is constructed 
by repeated application of the formula~\exref{Gn_rec_rln}.

Let us substitute the formula~\exref{Gn_rec_rln} with~$n=1$ 
for the first-order response function into the right-hand side 
of~\eqref{model_eq_Z}:
\bea
\nonumber
\dt Z(t) = \LL\int_0^t\diff t_1\,\kappa(t-t_1)Z(t_1)
\qquad\qquad\qquad\qquad\\ 
+\,\,T^{i\b_1}_{k\a_1} T^{m\b_2}_{n\a_2} \L^k_i \L^n_m
\int_0^t\diff t_1 \int_{t_1}^t\diff t' \int_0^{t'}\diff t_2\, 
\kappa(t-t_1)\kappa(t'-t_2) 
G^{\a_1\a_2}_{\b_1\b_2}(t'|t_1,t_2),
\label{Z_via_G2}
\eea 
where the operator~$\LL$ is defined in~\exref{def_L_iso}.
We now use the formula~\exref{Gn_rec_rln} to express 
the second-order response function on the right-hand side 
of the above equation: for~$t_2>t_1$, we have
\bea
\nonumber
G^{\a_1\a_2}_{\b_1\b_2}(t'|t_1,t_2) = 
\L^{\a_2}_{\b_2} G^{\a_1}_{\b_1}(t_2|t_1)
\qquad\qquad\qquad\\
+\,\,\L^q_p T^{p\b_3}_{q\a_3}
\int_{t_2}^{t'}\diff t'' \int_0^{t''}\diff t_3\,\kappa(t''-t_3)
G^{\a_1\a_2\a_3}_{\b_1\b_2\b_3}(t''|t_1,t_2,t_3), 
\label{G2_rec}
\eea
while for~$t_2<t_1$ we flip the variables, $t_1\leftrightarrow t_2$,
to make sure that the first-order response function on the right-hand 
side do not vanish: 
\bea
\nonumber
G^{\a_1\a_2}_{\b_1\b_2}(t'|t_1,t_2) = 
\L^{\a_1}_{\b_1} G^{\a_2}_{\b_2}(t_1|t_2)
\qquad\qquad\qquad\\
+\,\,\L^q_p T^{p\b_3}_{q\a_3}
\int_{t_1}^{t'}\diff t'' \int_0^{t''}\diff t_3\,\kappa(t''-t_3)
G^{\a_2\a_1\a_3}_{\b_2\b_1\b_3}(t''|t_2,t_1,t_3).  
\label{G2_rec_flip}
\eea
The recursion relation~\exref{Gn_rec_rln} is now applied 
to the first-order response functions in 
the formulas~\exref{G2_rec} and~\exref{G2_rec_flip}:
\bea
\label{G1_rec}
G^{\a_1}_{\b_1}(t_2|t_1) = \L^{\a_1}_{\b_1}Z(t_1)
+ \L^q_p T^{p\b_3}_{q\a_3}
\int_{t_1}^{t_2}\diff t'' \int_0^{t''}\diff t_3\,\kappa(t''-t_3)
G^{\a_1\a_3}_{\b_1\b_3}(t''|t_1,t_3),\\
\label{G1_rec_flip}
G^{\a_2}_{\b_2}(t_1|t_2) = \L^{\a_2}_{\b_2}Z(t_2)
+ \L^q_p T^{p\b_3}_{q\a_3}
\int_{t_2}^{t_1}\diff t'' \int_0^{t''}\diff t_3\,\kappa(t''-t_3)
G^{\a_2\a_3}_{\b_2\b_3}(t''|t_2,t_3).\,
\eea
All this must be substituted into~\eqref{Z_via_G2}:
\bea
\nonumber
\dt Z(t) &=& \LL\int_0^t\diff t_1\,\kappa(t-t_1)Z(t_1)\\
\nonumber
&+& T^{i\b_1}_{k\a_1}T^{m\b_2}_{n\a_2}
\L^k_i\L^n_m\L^{\a_1}_{\b_1}\L^{\a_2}_{\b_2}
\int_0^t\diff t_1\int_{t_1}^t\diff t'\int_0^{t_1}\diff t_2\,
\kappa(t-t_1)\kappa(t'-t_2)Z(t_2)\\
\nonumber
&+& T^{i\b_1}_{k\a_1}T^{m\b_2}_{n\a_2}
\L^k_i\L^n_m\L^{\a_2}_{\b_2}\L^{\a_1}_{\b_1}
\int_0^t\diff t_1\int_{t_1}^t\diff t'\int_{t_1}^{t'}\diff t_2\,
\kappa(t-t_1)\kappa(t'-t_2)Z(t_1)\\
&+& R(t),
\label{Z_with_R}
\eea 
where the remainder~$R(t)$ contains the assembled terms 
that involve quintuple time integrals. 

So far, all the manipulations we have carried out have been exact. 
It is now not hard to perceive the emerging contours of 
the small-$\tcorr$ expansion. Since the time-correlation 
function~$\kappa(t-t')$ is a profile of width~$\sim\tcorr$, 
the area under which is constant and equal to~$\kbar$, 
the triple time integrals in~\eqref{Z_with_R} are of 
the order of~$\tcorr\kbar^2$, while the quintuple time 
integrals absorbed into~$R(t)$ are of the order of~$\tcorr^2\kbar^3$. 
Further application of the recursion formula~\exref{Gn_rec_rln} 
to the second- and third-order response functions 
in the equations~\exref{G2_rec}--\exref{G1_rec_flip} 
leads to the appearance of more multiple time integrals 
of the time-correlation function~$\kappa(t-t')$. 
These integrals are of orders~$\tcorr^2\kbar^3$, 
$\tcorr^3\kbar^4$,~etc. 
We would only like to keep terms up to first order in the 
correlation time. The remainder term~$R(t)$ in~\eqref{Z_with_R}
can therefore be dropped.\\ 

{\em Remark on the physics of the $\tau$~expansion.} 
The above argument is based on the stipulation made 
at the beginning of this Section that the $\tau$~expansion 
must be carried out keeping the integral of the velocity 
time correlation function constant [formula~\exref{kbar_const}]. 
This requirement is natural because it leads to finite 
dynamo growth rates in the limit of zero correlation 
time [see~\ssecref{KT_method_Mark}, \eqref{expected_outcome}]. 
However, it is also acceptable to institute an alternative, 
arguably more physical, requirement that the {\em total energy} 
of the velocity field (i.e., the rms velocity) 
remain constant (as, e.g, in the numerics of Ref.~\cite{Kinney_etal_2D}). 
Quantitatively, this means that~$\kappa(0)$, 
rather than~$\int_0^\infty\diff\tau\,\kappa(\tau)$, 
is kept fixed. 
Under this constraint, the terms that 
we have previously estimated to be of orders~$\kbar$, 
$\tcorr\kbar^2$, $\tcorr^2\kbar^3$,~etc., and hence, $\kbar$ being 
constant, to represent the zeroth, first, second,~etc.~orders of the 
$\tau$~expansion, should now be reevaluated as follows. 
Since $\kbar\sim\tcorr\kappa(0)$, these terms are 
of orders~$\tcorr\kappa(0)$, $\tcorr^3\kappa(0)^2$, 
$\tcorr^5\kappa(0)^3$,~etc., and therefore constitute 
the first, third, fifth,~etc.~orders of the expansion.  
The shortcoming of this approach is that the dynamo growth 
rates vanish when~$\tcorr=0$, so formally there is no 
nontrivial zero-correlation-time limit.  
With $\kbar=\const$, this problem was avoided because the energy 
was formally infinite when~$\tcorr=0$ (a $\delta$-correlated 
velocity field cannot have a finite energy).
In any event, we see that, since the difference between 
keeping~$\kbar$ and~$\kappa(0)$ constant
does not affect the relative magnitudes  
of the terms in the expansion, our expansion scheme remains 
valid in both cases. Let us therefore proceed with our construction.\\
 
The dependence of the right-hand side of~\eqref{Z_with_R} 
on the ``past'' values of~$Z$ (i.e., on its values at times 
preceding~$t$) can also be resolved in the framework of 
the small-$\tcorr$ expansion. Formally integrating~\eqref{Z_with_R}, 
we get, at times~$t_1<t$, 
\bea
Z(t_1) = Z(t) - \LL\int_{t_1}^{t}\diff t'\int_0^{t'}\diff t_2\,
\kappa(t'-t_2)Z(t_2) + \cdots
\eea 
Upon substituting this onto the right-hand side of~\eqref{Z_with_R} 
and again discarding all the terms of orders higher than 
the first in~$\tcorr$, we~get
\bea
\nonumber
\dt Z(t) &=& \LL\int_0^t\diff t_1\,\kappa(t-t_1)Z(t)\\
\nonumber
&-& \LL^2\int_0^t\diff t_1\int_{t_1}^t\diff t'\int_0^{t'}\diff t_2\,
\kappa(t-t_1)\kappa(t'-t_2)Z(t)\\
\nonumber
&+& (\LL^2-\LL_1)
\int_0^t\diff t_1\int_{t_1}^t\diff t'\int_0^{t_1}\diff t_2\,
\kappa(t-t_1)\kappa(t'-t_2)Z(t)\\
&+& (\LL^2-\LL_2)
\int_0^t\diff t_1\int_{t_1}^t\diff t'\int_{t_1}^{t'}\diff t_2\,
\kappa(t-t_1)\kappa(t'-t_2)Z(t).
\label{Z_expanded}
\eea 
We have introduced the following two operators:
\bea
\label{def_L1}
\LL_1 &=& \L^k_i T^{i\b_1}_{k\a_1} \bigl[\L^{\a_1}_{\b_1},\L^n_m\bigr]      
T^{m\b_2}_{n\a_2}\L^{\a_2}_{\b_2} 
= {d^2\over d+1}\,\LL,\\
\label{def_L2}
\LL_2 &=& \L^k_i T^{i\b_1}_{k\a_1} \bigl[\L^{\a_1}_{\b_1},\LL\bigr] 
= 2d\LL,
\eea
where the square brackets denote commutators. 
We see that the terms in~\eqref{Z_expanded} that contain~$\LL^2$ 
cancel out, and only those terms remain that are due to 
the non-self-commuting nature of the operator~$\L^k_i$. 

Finally, we inverse-Fourier transform~\eqref{Z_expanded} 
into the $\vB$~space and take the long-time limit, $t\gg\tcorr$. 
The following Fokker--Planck equation with constant coefficients 
results:
\bea
\label{model_FPEq_1}
\dt P = {\kbar\over2}\[1 - \tcorr\kbar d
\({1\over2}{d\over d+1}\,K_1 + K_2\)\] \LL P,
\eea 
where the coefficients, 
\bea
\label{def_K1}
K_1 &=& {4\over\tcorr\kbar^2}\lim_{t\to\infty}
\int_0^t\diff t_1\int_{t_1}^t\diff t_2\int_0^{t_1}\diff t_3\,
\kappa(t-t_1)\kappa(t_2-t_3),\\
\label{def_K2}
K_2 &=& {4\over\tcorr\kbar^2}\lim_{t\to\infty}
\int_0^t\diff t_1\int_{t_1}^t\diff t_2\int_{t_1}^{t_2}\diff t_3\,
\kappa(t-t_1)\kappa(t_2-t_3),
\eea
are constants that depend on the particular shape of 
the time-correlation function~$\kappa(t-t')$. 
Thus, the $\tau$~expansion is {\em nonuniversal} in the sense 
that the specific choice of the functional form of 
the small-time regularization directly affects the values 
of the expansion coefficients (cf.~Ref.~\cite{Boldyrev_tcorr}). 
As an example, let us give the values of the coefficients~$K_1$ 
and~$K_2$ for two popular choices of~$\kappa(t-t')$: 
\bea
\kappa(t-t') = 
{\kbar\over2\tcorr}\,\exp\bigl[-|t-t'|/\tcorr\bigr]
&\Longrightarrow& K_1 = K_2 = 0.5\\
\kappa(t-t') = 
{\kbar\over\sqrt{\pi}\tcorr}\,\exp\bigl[-(t-t')^2/\tcorr^2\bigr]
&\Longrightarrow& K_1 \approx 0.33, \quad K_2 \approx 0.23.
\eea

In \secref{sec_KT_dynamo}, we will apply the method we have 
presented above to the more realistic general compressible kinematic 
dynamo problem in the Eulerian frame.\\

{\em Remark on the Kliatskin--Tatarskii method.}
The Gaussian hierarchy given by the equations~\exref{Gn_eq} 
and~\exref{Gn_tt} and based on repeated application of the 
Furutsu--Novikov formula was proposed by Kliatskin and 
Tatarskii~\cite{Kliatskin_Tatarskii} as a basis for constructing 
successive-approximation solutions of the problem of light propagation 
in a medium with randomly distributed inhomogeneities.  
Their method in its original form was carried over to the mean-field 
dynamo theory with finite-time-correlated velocity field 
by Vainshtein~\cite{Vainshtein_KT}. 
The method we have outlined in this section, while also 
based on the response-function hierarchy~\exref{Gn_eq}--\exref{Gn_tt}, 
differs substantially from that developed and applied 
by these authors. Their successive-approximation scheme 
consisted essentially in writing out the first $n$~equations 
in the hierarchy~\exref{Gn_eq}--\exref{Gn_tt} and then truncating 
it at the $n$th step by replacing~$\kappa(t-t_{n+1})$ by 
$\kbar\delta(t-t_{n+1})$ 
in the equation for the $n$th-order response function. 
This gave a closed system of equations that could be solved. 
Carried out in the first order, such a procedure would 
correspond to setting~$\kappa(t'-t_2)=\kbar\delta(t'-t_2)$ 
in~\eqref{Z_via_G2} and 
consequently~$\kappa(t_2-t_3)=\kbar\delta(t_2-t_3)$ 
in the expressions~\exref{def_K1} and~\exref{def_K2} 
for the coefficients~$K_1$ and~$K_2$. Such a substitution 
leads to~$K_1=0$ 
and~$K_2=(2/\tcorr\kbar)\int_0^\infty\diff\tau\,\tau\kappa(\tau)$, 
which is incorrect. 
The reason for this discrepancy is that, in the time integrals 
involving multiple products of the correlation 
functions~$\kappa(t-t_1)$, $\kappa(t_2-t_3)$, etc., 
the latter cannot be approximated by $\delta$~functions plus 
first-order corrections even in the small-$\tcorr$ limit.

\subsection{Comparison with the Van Kampen Cumulant Expansion Method}
\label{VK_method}

The evolution equation~$\exref{model_eq_Ztilde}$ 
for the ``unaveraged characteristic function''~$\tZ(t;\mu)$ 
is a stochastic linear differential equation whose form agrees 
exactly with that of the general such equation considered by 
van~Kampen~\cite{vanKampen,vanKampen_review} 
and simultaneously by Terwiel~\cite{Terwiel}:
\bea
\dt\tZ(t) = \A(t)\tZ(t),
\eea 
where, in our case, $\A(t) = \L^k_i\sigma^i_k(t)$.
In his work, van~Kampen developed a formalism that allowed one 
to construct successive terms in the short-correlation-time 
expansion of~$Z=\langle\tZ\rangle$ in terms of the cumulants 
of the operator~$\A$. Terwiel's projection-operator method 
was shown by its author to be equivalent to that of van~Kampen. 

Let us see what happens if the passive-advection 
problem given by~\eqref{model_B}, or, equivalently, 
by~\eqref{model_eq_Ztilde} is subjected to van~Kampen's 
expansion algorithm. The latter proceeds as follows.

Start by writing the formal solution 
of~\eqref{model_eq_Ztilde} in terms of the time-ordered 
exponential:
\bea
\nonumber
\tZ(t) = \l\lceil\exp\int_0^t\diff t'\A(t')\r\rceil \tZ(0)
\qquad\qquad\qquad\qquad\\
=\,\[1+ \int_0^t\diff t_1\,\A(t_1) + 
\int_0^t\diff t_1\int_0^{t_1}\diff t_2\,\A(t_1)\A(t_2) + 
\cdots\]\tZ(0).
\eea
This solution is averaged assuming that the initial 
distribution of~$\tZ$ is independent of the statistics of~$\A$:
\bea
\nonumber
Z(t) = \biggl[1 + 
\int_0^t\diff t_1\int_0^{t_1}\diff t_2\,
\bigl<\A(t_1)\A(t_2)\bigr>\biggr.
\qquad\qquad\qquad\qquad\\ 
+\,\biggl.\int_0^t\diff t_1\int_0^{t_1}\diff t_2
\int_0^{t_2}\diff t_3\int_0^{t_3}\diff t_4\,
\bigl<\A(t_1)\A(t_2)\A(t_3)\A(t_4)\bigr> + \cdots\biggr]Z(0).
\label{Z_via_Z0}
\eea 
Here all the odd-order averages have vanished 
(recall that~$\A = \L^k_i\sigma^i_k$).
The closed equation for~$Z(t)$ is now obtained as follows. 
First, the formal solution~\exref{Z_via_Z0} is differentiated 
with respect to time:
\bea
\nonumber
\dt Z(t) = \biggl[ 
\int_0^t\diff t_1\,\bigl<\A(t)\A(t_1)\bigr>\biggr.
\qquad\qquad\qquad\qquad\\ 
\biggl.+\,\int_0^t\diff t_1\int_0^{t_1}\diff t_2\int_0^{t_2}\diff t_3\,
\bigl<\A(t)\A(t_1)\A(t_2)\A(t_3)\bigr> + \cdots\biggr]Z(0).
\label{dtZ_via_Z0}
\eea
Second, $Z(0)$~is expressed in terms of~$Z(t)$ by formally 
inverting the operator series on the right-hand side of~\eqref{Z_via_Z0}, 
whereupon $Z(0)$~is substituted into~\eqref{dtZ_via_Z0}. 
Keeping only the terms that contain up to three time integrations, 
as we did in the previous section, we~get
\bea
\nonumber
\dt Z(t) &=& \biggl[ 
\int_0^t\diff t_1\,\bigl<\A(t)\A(t_1)\bigr>\biggr.\\
\nonumber
&+&\int_0^t\diff t_1\int_0^{t_1}\diff t_2\int_0^{t_2}\diff t_3\,
\bigl<\A(t)\A(t_1)\A(t_2)\A(t_3)\bigr>\\
&-&\biggl.\int_0^t\diff t_1\int_0^t\diff t_2\int_0^{t_2}\diff t_3\,
\bigl<\A(t)\A(t_1)\bigr>\bigl<\A(t_2)\A(t_3)\bigr> + \cdots\biggr] Z(t).
\label{Z_eq_A}
\eea
The quadruple average in the above expression splits into 
three products of second-order averages in the usual Gaussian way. 
Since 
\bea
\label{A_avg_2}
\bigl<\A(t)\A(t_1)\bigr> = \kappa(t-t_1)\,T^{ij}_{kl}\L^k_i\L^l_j 
= \kappa(t-t_1)\,\LL,
\eea
we have
\bea
\nonumber
\bigl<\A(t)\A(t_1)\A(t_2)\A(t_3)\bigr> 
&=& \kappa(t-t_1)\kappa(t_2-t_3)\, 
T^{ij}_{kl} T^{mn}_{pq}\L^k_i\L^l_j\L^p_m\L^q_n\\ 
\nonumber
&+&\kappa(t-t_2)\kappa(t_1-t_3)\, 
T^{im}_{kp} T^{jn}_{lq}\L^k_i\L^l_j\L^p_m\L^q_n\\ 
\nonumber
&+&\kappa(t-t_3)\kappa(t_1-t_2)\, 
T^{in}_{kq} T^{jm}_{lp}\L^k_i\L^l_j\L^p_m\L^q_n\\
\nonumber
&=& \kappa(t-t_1)\kappa(t_2-t_3)\,\LL^2\\
\nonumber
&+& \kappa(t-t_2)\kappa(t_1-t_3)\,\bigl(\LL^2 - \LL_1\bigr)\\
&+& \kappa(t-t_3)\kappa(t_1-t_2)\,\bigl(\LL^2 - \LL_2\bigr),
\label{A_avg_4}\\
\bigl<\A(t)\A(t_1)\bigr>\bigl<\A(t_2)\A(t_3)\bigr> 
&=& \kappa(t-t_1)\kappa(t_2-t_3)\,\LL^2. 
\label{A_avg_22}
\eea 
The operators~$\LL$, $\LL_1$, and~$\LL_2$ are the same as those 
in the previous section [see definitions~\exref{def_L_iso}, 
\exref{def_L1}, and~\exref{def_L2}].
The averages~\exref{A_avg_2}, \exref{A_avg_4}, and~\exref{A_avg_22} 
are now substituted into~\eqref{Z_eq_A}. 
The triple time integrals can be argued to represent 
(all of the) first-order terms in the small-$\tcorr$ expansion 
in the same way as it was done in~\ssecref{KT_method_exp}. 
In the limit~$t\gg\tcorr$, the coefficients in~\eqref{Z_eq_A} 
do not depend on time. The resulting Fokker--Planck equation 
for the PDF~$P(t;B)$ 
[which is the inverse Fourier transform of~$Z(t;\mu)$]
obtained by the van~Kampen method and analogous 
to~\eqref{model_FPEq_1} is then
\bea
\label{FPEq_vKampen}
\dt P = {\kbar\over2}\[1 - \tcorr\kbar d
\({1\over2}{d\over d+1}\,C_1 + C_2\) 
- {1\over2}\,\tcorr\kbar\bigl(C_0-C_1-C_2\bigr)\LL\] \LL P,
\eea
where the coefficients~$C_0$, $C_1$, and~$C_2$, that depend 
on the shape function~$\kappa(t-t')$, are as follows:
\bea
\label{def_C0}
C_0 &=& {4\over\tcorr\kbar^2}\lim_{t\to\infty}
\int_0^t\diff t_1\int_{t_1}^t\diff t_2\int_0^{t_2}\diff t_3\,
\kappa(t-t_1)\kappa(t_2-t_3),\\
\label{def_C1}
C_1 &=& {4\over\tcorr\kbar^2}\lim_{t\to\infty}
\int_0^t\diff t_1\int_0^{t_1}\diff t_2\int_0^{t_2}\diff t_3\,
\kappa(t-t_2)\kappa(t_1-t_3),\\
\label{def_C2}
C_2 &=& {4\over\tcorr\kbar^2}\lim_{t\to\infty}
\int_0^t\diff t_1\int_0^{t_1}\diff t_2\int_0^{t_2}\diff t_3\,
\kappa(t-t_3)\kappa(t_1-t_2).
\eea
By comparing the definition of~$C_0$ with those of 
the coefficients~$K_1$ and~$K_2$ in~\ssecref{KT_method_exp} 
[see formulas~\exref{def_K1} and~\exref{def_K2}], 
we immediately establish that $C_0=K_1+K_2$. 
Furthermore, it is also not hard to ascertain 
that~$C_1=K_1$ and~$C_2=K_2$. Therefore, the last term 
in~\eqref{FPEq_vKampen} vanishes, and the first-order 
Fokker--Planck equations~\exref{model_FPEq_1} and~\exref{FPEq_vKampen} 
are identical.
Thus, the results obtained via the van~Kampen method 
are consistent with ours.
Unlike the van~Kampen method, 
however, our method does not involve any nontrivial operator 
algebra and is therefore better suited for a wide variety of applications. 
In particular, the stochastic equations containing spatial 
derivatives (such as the convective derivatives present in all Eulerian 
passive-advection problems) can be handled without much 
additional difficulty (this will be done in detail for the full 
kinematic dynamo problem in~\secref{sec_KT_dynamo}).

\subsection{Discussion: The Vertex Corrections}
\label{tau_discussion}

\begin{figure} 
\centerline{\psfig{file=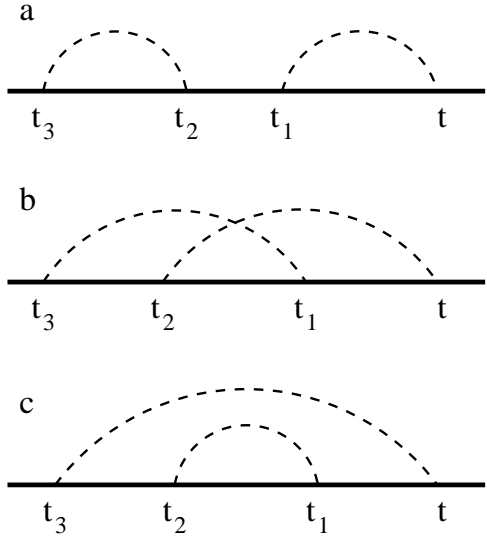}}
\vskip0.25in
\caption{\small The three possible fourth-order diagrams. 
The times~$t_1$, $t_2$, $t_3$ are integration variables 
and may float along the axis, but their positions relative 
to each other and to the time~$t$ are determined by the limits 
of integration and therefore preserved. 
A dashed line connecting any two times~$t_i$ and~$t_j$ 
represents the time-correlation function~$\kappa(t_i-t_j)$ 
in the integrand of a triple time-history integral.}
\label{fig_diagrams}
\end{figure}

While the particular methods one employs to obtain the successive 
terms in the $\tau$~expansion may vary and depend on one's taste 
and the specific demands of the stochastic problem at hand, 
the underlying structure of the $\tau$~expansion remains the same 
and is rooted in the common properties of all turbulence closure 
problems (see, e.g.,~Ref.~\cite{Krommes_review}). As we have stated 
in general terms in the introduction to this paper, and 
as was clear from our construction of the response-function formalism 
in~\ssecref{KT_method_hierarchy}--\ssecref{KT_method_exp} 
or of van~Kampen's explicit series solution~\exref{Z_via_Z0} 
in~\ssecref{VK_method}, 
averaged solutions of stochastic equations such 
as~\eqref{model_eq_Ztilde} can be represented in terms 
of infinite sums of multiple time-history integrals containing 
products of time-correlation functions~$\kappa(t_i-t_j)$ in 
the integrands. This summation can be visualized in terms 
of Feynman-style diagrams. The $n$-point 
diagrams represent the terms containing $n$~time-history 
integrations. As an example, \figref{fig_diagrams} lists the three 
possible four-point diagrams.  

It was noted by Kazantsev~\cite{Kazantsev} (see also 
Ref.~\cite{Vainshtein_Zeldovich}) 
that the white-noise approximation corresponds to the partial 
summation of all ladder-type diagrams such as the four-point 
one shown in \figref{fig_diagrams}(a).
The distinctive property of these diagrams is that the pairs 
of points $t_i$,~$t_j$ 
at which the time-correlation functions in the integrands 
of the time-history integrals are taken, can be fused without  
interfering with each other. No essential information is therefore 
lost when the time-correlation functions~$\kappa(t_i-t_j)$ are 
approximated by $\delta$~functions. However, in all orders of 
the $\tau$~expansion but the zeroth, diagrams with more 
tangled topology appear: e.g., in the first order, these are the 
diagrams~\ref{fig_diagrams}(b) and~\ref{fig_diagrams}(c)]. 
Such diagrams are 
often referred to as the vertex corrections. Fusing points in 
these diagrams leads to the loss of terms that cannot 
be neglected~\cite{fnote_KT_fusing_pts}.
This is the context in which 
the emerging nonuniversality with respect to the shape of the 
time-correlation profile should be viewed. 

In this paper, we restrict our consideration to the first-order 
terms in the $\tau$~expansion. The relevant diagrams are 
the four-point ones shown in~\figref{fig_diagrams}. 
The diagrams~\ref{fig_diagrams}(b) and~\ref{fig_diagrams}(c) give 
rise to the coefficients~$C_1$ and~$C_2$, respectively 
[see~\eqref{FPEq_vKampen} and formulas~\exref{def_C1} 
and~\exref{def_C2}]. Upon changing variables $t_1\leftrightarrow t_2$ 
in the diagram~\ref{fig_diagrams}(b) and 
$t_1\rightarrow t_2$, $t_2\rightarrow t_3$, $t_3\rightarrow t_1$ 
in the diagram~\ref{fig_diagrams}(c), we see that these diagrams 
equally well correspond to the coefficients~$K_1$ and~$K_2$ 
[\eqref{model_FPEq_1} and formulas~\exref{def_K1} 
and~\exref{def_K2}]. 

\section{The Functional Expansion for the Kinematic Dynamo 
in a Finite-Time-Correlated Velocity Field}
\label{sec_KT_dynamo}

In this Section, we use the functional expansion method 
developed in~\secref{sec_tau_exp} to construct the $\tau$~expansion 
for the general diffusion-free
kinematic dynamo problem in the Eulerian frame 
with an arbitrarily compressible velocity field. 

Through the convective derivative, an explicit spatial 
dependence is now present in the problem. This leads 
to the appearance of the new effect advertised in the Introduction:  
while the zeroth-order terms 
in the expansion only depend on the one-point correlation 
properties of the velocity gradients, 
the first-order terms also depend on 
the {\em energy} of the advecting velocity field and on the one-point 
correlation function of its {\em second} derivatives. 
The former represents the loss of Galilean invariance, the latter 
the loss of the small-scale universality and the advent of the 
sensitive dependence of the statistics on the large-scale 
structure of the velocity correlations.  

In this Section, all statistics are {\em Eulerian.} 
For the questions regarding the transformation of PDF's of passive 
fields from the Eulerian to the Lagrangian frame, we address the reader 
to Ref.~\cite{BS_metric}, 
as well as to Ref.~\cite{Boldyrev_tcorr}, 
where the $\tau$~expansion is treated as a problem in stochastic 
calculus and Lagrangian statistics are discussed.

\subsection{The Gaussian Hierarchy}
\label{hierarchy}

The magnetic field passively advected by the 
velocity field~$\xi^i(t,\vx)$ evolves according to 
the Hertz induction equation (formally in $d$~dimensions):
\bea
\label{ind_eq}
\dt B^i = - \xi^k B^i_{,k} + \xi^i_{,k} B^k - \xi^k_{,k} B^i,
\eea
where $\xi^i_{,k}=\d\xi^i/\d x^k$, $B^i_{,k}=\d B^i/\d x^k$, and 
the Einstein summation convention is used throughout. 
Let the advecting velocity field $\xi^i(t,\vx)$ be a homogeneous 
and isotropic 
Gaussian random field whose statistics are defined by its 
second-order correlation tensor: 
\bea
\bigl<\xi^i(t,\vx)\xi^j(t',\vx')\bigr> = \kappa^{ij}(t-t',\vx-\vx'),
\eea
where, as a function of the time separation~$t-t'$, 
the correlator~$\kappa^{ij}$ is 
assumed to have a finite width~$\tcorr$, which we will call 
{\em the velocity correlation time.} 
As we will only study the one-point statistics of the magnetic field, 
all relevant information about the velocity correlation properties 
is contained in the Taylor expansion of~$\kappa^{ij}$ 
around the origin:
\bea
\nonumber
\kappa^{ij}(\tau,\vy)\,\,=\,\,\kappa_0(\tau)\delta^{ij} 
&-& {1\over 2}\,\kappa_2(\tau)\bigl[y^2\delta^{ij} + 2a y^iy^j\bigr]\\
\label{xi_024} 
&+&{1\over 4}\,\kappa_4(\tau)y^2\bigl[y^2\delta^{ij} + 2b y^iy^j\bigr] 
+ \cdots,
\eea 
as~$y\to0$. 
Here~$a$ and~$b$ are the compressibility parameters. 
Between the purely incompressible and the purely irrotational cases, 
they vary in the intervals
\bea
-{1\over d+1}\le a \le 1, \qquad -{2\over d+3}\le b \le 2.
\eea
We should like to mention here that the choice of the coefficients 
of the small-scale expansion~\exref{xi_024} of the velocity correlation 
tensor is, strictly speaking, not entirely unconstrained. 
As~$\kappa^{ij}(\tau,\vy)$ is a {\em correlation function}, 
it must be an inverse Fourier transform of a proper correlation 
function in the Fourier space~\cite{Monin_Yaglom}. 
In~\Apref{ap_k_space}, we give the expressions for 
the coefficients of the expansion~\exref{xi_024} 
in terms of the spectral characteristics of the velocity field. 
We further note that, while $a$~and $b$~can certainly 
be functions of~$\tau$, 
we will not overly shrink the limits of physical generality by 
assuming that they are either constant or slowly-varying functions 
of time, i.e.~that they do not change appreciably over one correlation 
time. 

In order to determine the one-point statistics of the magnetic field, 
we follow the standard procedure~\cite{Polyakov} and introduce the 
characteristic function of~$B^i(t,\vx)$ at an arbitrary fixed spatial 
point~$\vx$:
\bea
Z(t;\mu) = \bigl<\tZ(t,\vx;\mu)\bigr> 
= \bigl<\exp\bigl[i\mu_i B^i(t,\vx)\bigr]\bigr>. 
\eea
As usual, the angle brackets denote ensemble 
averages and the overtildes mark unaveraged quantities. 
The function~$Z$ is the Fourier transform of the~PDF of the vector 
elements~$B^i$. Due to spatial homogeneity,
$Z$~does not depend on the point~$\vx$, where~$B^i(t,\vx)$ is taken. 
Upon differentiating~$\tZ$ with respect to time and using~\eqref{ind_eq}, 
we get
\bea
\label{Ztilde_eq}
\dt\tZ = -\xi^k\tZ_{,k}
+ \xi^i_{,k}\mu_i{\d\over\d\mu_k}\tZ 
- \xi^k_{,k}\mu_l{\d\over\d\mu_l}\tZ 
= -\xi^k\tZ_{,k} + \L^k_i\,\xi^i_{,k}\tZ,
\eea
where, for the sake of future convenience, we introduce 
the operator
\bea
\label{def_Lambda}
\L^k_i = \mu_i{\d\over\d\mu_k} - \delta^k_i\mu_l{\d\over\d\mu_l},
\eea
which will turn up repeatedly in this calculation.

In order to obtain an evolution equation for the characteristic 
function~$Z(t;\mu)$ of the random magnetic field, 
we average both sides of~\eqref{Ztilde_eq}. 
Since, due to the homogeneity of the problem,
$\langle\xi^k\tZ_{,k}\rangle = - \langle\xi^k_{,k}\tZ\rangle$, 
we may write the equation for~$Z(t;\mu)$ in the following form:
\bea
\nonumber
\dt Z(t) = \bigl(\delta^k_i + \L^k_i\bigr)
\bigl<\xi^i_{,k}(t,\vx)\tZ(t,\vx)\bigr>
\qquad\qquad\qquad\\ 
= -\bigl(\delta^k_i + \L^k_i\bigr)
\int_0^t\diff t_1\intx_1\,
\kappa^{i\b_1}_{,k\a_1}(t-t_1,\vx-\vx_1)
G^{\a_1}_{\b_1}(t,\vx|t_1,\vx_1),
\label{Z_eq}
\eea
where the mixed average on the right-hand side has been ``split''
with the aid of the Furutsu--Novikov (``Gaussian-integration'') 
formula~\cite{Furutsu,Novikov}, 
and the $\mu$~dependence in the arguments has been suppressed 
for the sake of notational compactness.
We have introduced the first-order averaged response function 
of the following species:
\bea 
G^{\a_1}_{\b_1}(t,\vx|t_1,\vx_1) = 
\bigl<\tG^{\a_1}_{\b_1}(t,\vx|t_1,\vx_1)\bigr>
= \<{\delta\tZ(t,\vx)\over\delta\xi^{\b_1}_{,\a_1}(t_1,\vx_1)}\>. 
\eea
As a response function, $G^{\a_1}_{\b_1}$~satisfies the causality 
constraint:~$G^{\a_1}_{\b_1}(t,\vx|t_1,\vx_1) = 0$ for~$t_1>t$.
The same-time form of~$G^{\a_1}_{\b_1}$ can be obtained in terms 
of the characteristic function~$Z(t)$:
integrating~\eqref{Ztilde_eq} from~$0$ to~$t_1$, taking the functional 
derivative~$\delta/\delta\xi^{\b_1}_{,\a_1}(t',\vx_1)$, 
averaging, setting~$t'=t_1$, and taking causality into account, 
we~get 
\bea
\label{G_same}
G^{\a_1}_{\b_1}(t_1,\vx|t_1,\vx_1) = 
\delta(\vx-\vx_1)\,\L^{\a_1}_{\b_1} Z(t_1).
\eea
In order to find~$G^{\a_1}_{\b_1}(t,\vx|t_1,\vx_1)$ at $t>t_1$, 
we take the functional derivative~$\delta/\delta\xi^{\b_1}_{,\a_1}(t',\vx')$ 
of both sides of~\eqref{Ztilde_eq} and establish that each element of 
the unaveraged tensor~$\tG^{\a_1}_{\b_1}$ satisfies 
an equation identical in form to~\eqref{Ztilde_eq}: 
\bea
\label{Gtilde_eq}
\dt\tG^{\a_1}_{\b_1}(t,\vx|t_1,\vx_1) = 
- \xi^m(t,\vx)\tG^{\a_1}_{\b_1,m}(t,\vx|t_1,\vx_1)
+ \L^n_m\xi^m_{,n}(t,\vx)\tG^{\a_1}_{\b_1}(t,\vx|t_1,\vx_1).
\eea
Subscripts such as ``$_{,m}$'' in the above equation mean, in accordance 
with the usual notation, the partial differentiation with respect 
to~$x^m$, viz.,~$\d/\d x^m$.

We must now average~\eqref{Gtilde_eq} in its turn, to obtain an evolution 
equation for~$G^{\a_1}_{\b_1}(t,\vx|t_1,\vx_1)$ at $t>t_1$. 
Using the initial condition~\exref{G_same}, let us write 
this evolution equation in the integral form valid for all~$t\ge t_1$:
\bea
\nonumber
G^{\a_1}_{\b_1}(t,\vx|t_1,\vx_1) = 
\delta(\vx-\vx_1)\,\L^{\a_1}_{\b_1} Z(t_1)\, 
\qquad\qquad\qquad\\ 
\nonumber 
- \int_{t_1}^t\diff t'\int_0^{t'}\diff t_2\intx_2
\Bigl[\kappa^{m\b_2}(t'-t_2,\vx-\vx_2) 
G^{\a_1}_{\b_1\b_2,m}(t',\vx|t_1,\vx_1;t_2,\vx_2)\Bigr.\\
+\,\Bigl.\kappa^{m\b_2}_{,n\a_2}(t'-t_2,\vx-\vx_2) 
G^{\a_1\a_2}_{\b_1\b_2}(t',\vx|t_1,\vx_1;t_2,\vx_2)\Bigr],
\qquad\qquad 
\label{G_eq}
\eea
where the mixed averages have again been ``split'' 
by the Furutsu--Novikov formula, at the price of introducing 
two new second-order response functions:
\bea
G^{\a_1}_{\b_1\b_2}(t',\vx|t_1,\vx_1;t_2,\vx_2) 
&=& \<\delta^2\tZ(t',\vx)\over
\delta\xi^{\b_1}_{,\a_1}(t_1,\vx_1)\delta\xi^{\b_2}(t_2,\vx_2)\>,\\
G^{\a_1\a_2}_{\b_1\b_2}(t',\vx|t_1,\vx_1;t_2,\vx_2)
&=& \<\delta^2\tZ(t',\vx)\over
\delta\xi^{\b_1}_{,\a_1}(t_1,\vx_1)\delta\xi^{\b_2}_{,\a_2}(t_2,\vx_2)\>.
\eea

In the same way that the equal-time first-order response function 
was expressed in terms of~$Z(t)$ [\eqref{G_same}],
the second-order response functions
at~$t'=t_1$ or~$t'=t_2$ can be expressed 
in terms of~$G^{\a_2}_{\b_2}(t_1,\vx|t_2,\vx_2)$ 
or~$G^{\a_1}_{\b_1}(t_2,\vx|t_1,\vx_1)$, respectively. 
Because of causality, 
the former representation would be valid provided~$t_1\ge t_2$, 
the latter in the opposite case~$t_2\ge t_1$.  
At other times, $t_1,t_2\le t'$, the functions 
$G^{\a_1}_{\b_1\b_2}$ and $G^{\a_1\a_2}_{\b_1\b_2}$ 
satisfy integral equations analogous to~\eqref{G_eq}, 
where third-order response functions make their appearance.
An infinite open hierarchy can thus be obtained by further iterating this 
procedure and introducing response functions of ascending orders.  
This hierarchy constitutes the exact description of the 
statistics of the kinematic dynamo problem with arbitrary 
velocity correlation time.

\subsection{The $\tau$ Expansion}
\label{tau_expansion}

The expansion in small correlation time must be 
carried out in such a way that the time integral of the 
velocity correlator~$\kappa^{ij}(\tau,\vy)$ remains 
constant.
Since $\kappa^{ij}(\tau,\vy)$~has 
a finite (small) width~$\tcorr$, we can conclude 
that the double time integral on the right-hand side of~\eqref{G_eq} 
must be of first order in the correlation time~$\tcorr$.  
As we are only interested in constructing the $\tau$~expansion 
up to first order, it is now sufficient to calculate the second-order 
response functions $G^{\a_1}_{\b_1\b_2}$ and $G^{\a_1\a_2}_{\b_1\b_2}$ 
with zeroth-order precision.   

We have already mentioned that recursive relations completely 
analogous to the relation~\exref{G_eq} can be derived for the second-order 
response functions. The latter are thereby expressed as their 
equal-time values plus double time integrals of the same sort 
as that which appeared on the right-hand side of~\eqref{G_eq}.
These time integrals are first order in the correlation time 
and can therefore be neglected. The equal-time 
values of the second-order response functions are obtained 
by formally integrating~\eqref{Ztilde_eq}, taking functional 
derivatives of it, averaging, and using causality. 
The second-order response functions are thus expressed 
to zeroth order in terms of the first-order ones. 
These latter can by the same token be replaced by their 
equal-time values, which only contain the characteristic 
function~$Z(t)$. The resulting expressions, valid to zeroth order, 
must be substituted into the first-order term (the double time 
integral) in~\eqref{G_eq}. All these manipulations, which require 
a fair amount of algebra, are relegated to~\apref{ap_response2}.
Here we simply give the resulting expression for the 
first-order response function, valid to first order in~$\tcorr$: 
\bea
\nonumber
G^{\a_1}_{\b_1}(t,\vx|t_1,\vx_1) = 
\delta(\vx-\vx_1)\biggl[\L^{\a_1}_{\b_1} Z(t_1)\biggr.
\qquad\qquad\qquad\\ 
\nonumber
-\,\,\int_{t_1}^t\diff t'\int_0^{t_1}\diff t_2\,
\kappa^{m\b_2}_{,n\a_2}(t'-t_2,0)
\bigl(\delta^n_m + L^n_m\bigr)\L^{\a_1}_{\b_1}\L^{\a_2}_{\b_2} Z(t_2)\qquad\\
\nonumber
-\,\,\int_{t_1}^t\diff t'\int_{t_1}^{t'}\diff t_2\,
\kappa^{m\b_2}_{,n\a_2}(t'-t_2,0)
\bigl(\delta^n_m + L^n_m\bigr)\L^{\a_2}_{\b_2}\L^{\a_1}_{\b_1} Z(t_1)\qquad\\
\nonumber
\biggl.+\,\,\int_{t_1}^t\diff t'\int_0^{t_1}\diff t_2\,
\kappa^{\a_1\b_2}_{,\b_1\a_2}(t'-t_2,0)\,
\L^{\a_2}_{\b_2} Z(t_2)\biggr]\qquad\qquad\\
+\,\,{\d^2\delta(\vx-\vx_1)\over\d x^m\d x^n}
\int_{t_1}^t\diff t'\int_{t_1}^{t'}\diff t_2\,
\kappa^{mn}(t'-t_2,0)\,\L^{\a_1}_{\b_1} Z(t_1) + {\cal O}(\tcorr^2).
\label{G_eq_1}
\eea

This expression must now be substituted into the time-history 
integral on the right-hand side of~\eqref{Z_eq}. 
This gives a closed integro-differential equation for the 
characteristic function~$Z(t)$. However, the dependence on 
the past values of~$Z$ is spurious and can be resolved 
to first order in~$\tcorr$. Indeed, we can formally 
integrate~\eqref{Z_eq} from~$t_1$ to~$t$ and, using the 
zeroth-order value of the first-order response function [the first 
term in the formula~\exref{G_eq_1}],~get
\bea
\label{Z_eq_1}
Z(t_1) = Z(t) + 
\bigl(\delta^n_m + \L^n_m\bigr)
\int_{t_1}^t\diff t'\int_0^{t'}\diff t_2\,
\kappa^{m\b_2}_{,n\a_2}(t'-t_2,0)\,
\L^{\a_2}_{\b_2} Z(t_2) + {\cal O}(\tcorr^2).
\eea
The double time integral in this equation is of first order in~$\tcorr$, 
as usual.  

Upon assembling the equations~\exref{Z_eq}, \exref{G_eq_1}, 
and~\exref{Z_eq_1}, we finally arrive at the following closed partial 
differential equation for~$Z(t)$:
\bea
\nonumber
\dt Z(t) = -\int_0^t\diff t_1\,
\kappa^{i\b_1}_{,k\a_1}(t-t_1,0)
\bigl(\delta^k_i + \L^k_i\bigr)\L^{\a_1}_{\b_1} Z(t)
\qquad\qquad\qquad\\
\nonumber
-\int_0^t\diff t_1\int_{t_1}^t\diff t'\int_0^{t_1}\diff t_2\,
\kappa^{i\b_1}_{,k\a_1}(t-t_1,0)
\qquad\qquad\qquad\qquad\qquad\qquad\qquad\quad\\
\nonumber
\times\,\kappa^{m\b_2}_{,n\a_2}(t'-t_2,0)
\bigl(\delta^k_i + \L^k_i\bigr)
\bigl[\L^{\a_1}_{\b_1},\L^n_m\bigr]\L^{\a_2}_{\b_2} Z(t)\qquad\\
\nonumber
-\int_0^t\diff t_1\int_{t_1}^t\diff t'\int_{t_1}^{t'}\diff t_2\,
\kappa^{i\b_1}_{,k\a_1}(t-t_1,0)
\qquad\qquad\qquad\qquad\qquad\qquad\qquad\quad\\
\nonumber
\times\,\kappa^{m\b_2}_{,n\a_2}(t'-t_2,0)
\bigl(\delta^k_i + \L^k_i\bigr)
\bigl[\L^{\a_1}_{\b_1},\bigl(\delta^n_m + \L^n_m\bigr)
\L^{\a_2}_{\b_2}\bigr]Z(t)\qquad\\
\nonumber
-\int_0^t\diff t_1\int_{t_1}^t\diff t'\int_0^{t_1}\diff t_2\,
\kappa^{i\b_1}_{,k\a_1}(t-t_1,0)\kappa^{\a_1\b_2}_{,\b_1\a_2}(t'-t_2,0)
\bigl(\delta^k_i + \L^k_i\bigr)\L^{\a_2}_{\b_2} Z(t)\qquad\\
-\int_0^t\diff t_1\int_{t_1}^t\diff t'\int_{t_1}^{t'}\diff t_2\,
\kappa^{i\b_1}_{,k\a_1mn}(t-t_1,0)\kappa^{mn}(t'-t_2,0)
\bigl(\delta^k_i + \L^k_i\bigr)\L^{\a_1}_{\b_1} Z(t).
\quad\,\,\,
\label{Z_eq_closed}
\eea 
The square brackets denote commutators.

Note that, besides the second derivatives of the velocity correlation 
tensor, the first-order terms contain the fourth ones, as well 
as the undifferentiated tensor itself [in the last term 
in~\eqref{Z_eq_closed}]. The latter implies the loss of Galilean 
invariance, the former the loss of small-scale universality in the 
sense that the large-scale structure of the velocity correlator 
starts to play a role. This effect could not have been captured 
if the velocity field had been assumed to be purely a combination 
of the instantaneous velocity at a given point and a linear shear.

\subsection{The Fokker--Planck Equation}
\label{Fokker_Planck_tcorr}

In order to obtain the Fokker--Planck equation for the PDF of 
the magnetic field, we must inverse-Fourier transform~\eqref{Z_eq_closed}  
back to $\vB$~dependence. The inverse Fourier transform of~$Z(t;\mu)$
is the one-point PDF~$P(t;\vB)$. 
We will continue using the symbol~$\L^k_i$ to denote 
the counterpart of the operator~$\L^k_i$ in the $\vB$~space: 
\bea
\label{Lambda_Bspace}
\L^k_i = (d-1)\,\delta^k_i - B^k{\d\over\d B^i} 
+ \delta^k_i B^l{\d\over\d B^l}.
\eea 
Due to the isotropy of the problem, the PDF~$P(t;\vB)$ will in fact 
be a scalar function of the field strength~$B$ only. Thus, all 
the operators that appear on the right-hand side of the $\vB$-space
counterpart of~\eqref{Z_eq_closed} must, after they are convolved 
with the velocity correlation tensors, be expressible in terms 
of~$B$.
Let us use the Taylor expansion~\exref{xi_024} of the velocity correlator 
to calculate the tensor convolutions in~\eqref{Z_eq_closed}. 
We have
\bea
\kappa^{ij}(\tau,0) &=& \kappa_0(\tau)\,\delta^{ij},\\
\kappa^{ij}_{,kl}(\tau,0) &=& 
-\kappa_2(\tau)\bigl[\delta^{ij}\delta_{kl} + 
a\,\bigl(\delta^i_k\delta^j_l + \delta^i_l\delta^j_k\bigr)\bigr] 
= -\kappa_2(\tau)\,T^{ij}_{kl},\\
\nonumber
\kappa^{ij}_{,klmm}(\tau,0) &=&
\kappa_4(\tau)\bigl[2\,(d+2+b)\,\delta^{ij}\delta_{kl} + 
(d+4)\,b\,\bigl(\delta^i_k\delta^j_l + 
\delta^i_l\delta^j_k\bigr)\bigr]\\
&=& \kappa_4(\tau)\,U^{ij}_{kl}.
\eea
A number of second-order differential operators (with respect to~$B$)
arise in~\eqref{Z_eq_closed}. In the zeroth-order term, 
we have
\bea
\label{deff_LL}
\LL = T^{i\b_1}_{k\a_1}\bigl(\delta^k_i + \L^k_i\bigr)\L^{\a_1}_{\b_1} 
= {d-1\over d+1}\,\biggl(B{\d\over\d B} + d\biggr)
\Biggl(\(1+\igamma\) B{\d\over\d B} + (d+1)\igamma\Biggr);
\eea
two operators appearing in the first-order terms result from 
the non-self-commuting nature of the operator~$\L^k_i$ 
[see the second and the third terms 
in~\eqref{Z_eq_closed}]~\cite{fnote_operators}:
\bea
\nonumber
\LL_1 &=& T^{i\b_1}_{k\a_1} T^{m\b_2}_{n\a_2}
\bigl(\delta^k_i + \L^k_i\bigr) 
\bigl[\L^{\a_1}_{\b_1},\L^n_m\bigr]\L^{\a_2}_{\b_2}\\
\label{deff_LL1}
&=& {d^2(d-1)\over(d+1)^2}\,\biggl(B{\d\over\d B} + d\biggr)
\biggl(1+{\igamma\over d^2}\biggr)^2 B{\d\over\d B},\\
\nonumber
\LL_2 &=& 
T^{i\b_1}_{k\a_1} T^{m\b_2}_{n\a_2}
\bigl(\delta^k_i + \L^k_i\bigr)
\bigl[\L^{\a_1}_{\b_1},\bigl(\delta^n_m+\L^n_m\bigr)\L^{\a_2}_{\b_2}\bigr] =
T^{i\b_1}_{k\a_1}\bigl(\delta^k_i + \L^k_i\bigr)
\bigl[\L^{\a_1}_{\b_1},\LL\bigr]\\
\label{deff_LL2}
&=& {2d(d-1)\over d+1}\,\biggl(B{\d\over\d B} + d\biggr)
\biggl(1+{\igamma\over d^2}\biggr) B{\d\over\d B};
\qquad
\eea
and, finally, there are two other operators due 
to the presence of the convective term (i.e., explicit spatial 
dependence) in the induction equation 
[see the fourth and the fifth terms in~\eqref{Z_eq_closed}]:
\bea 
\nonumber
\MM_1 &=& T^{i\b_1}_{k\a_1} T^{\a_1\b_2}_{\b_1\a_2}
\bigl(\delta^k_i + \L^k_i\bigr)\L^{\a_2}_{\b_2}\,=\, 
{d(d-1)\over(d+1)^2}\,\biggl(B{\d\over\d B} + d\biggr)\\
&&\qquad\times
\Biggl[\(1 + {2\igamma\over d^2} + {d(d+1)-1\over d^3}\,\igamma^2\)
B{\d\over\d B} + {(d+1)^2\over d^2}\,\igamma\Biggr],\quad
\label{deff_MM1}\\
\nonumber
\MM_2 &=& U^{i\b_1}_{k\a_1}
\bigl(\delta^k_i + \L^k_i\bigr)\L^{\a_1}_{\b_1}\,=\,
{2(d-1)(d+4)\over d+3}\,\biggl(B{\d\over\d B} + d\biggr)\\
&&\qquad\times
\Biggl[\(1 + {d^2+4d+2\over2d(d+4)}\,\zeta\)B{\d\over\d B} 
+ {(d+2)(d+3)\over2(d+4)}\,\zeta\Biggr].
\label{deff_MM2}
\eea
In all of the above, $\igamma = d\bigl[1+(d+1)a\bigr]$ and 
$\zeta = d\bigl[2+(d+3)b\bigr]$ are compressibility 
parameters that vanish in the case of incompressible flow.
In this latter case, the operators defined above simplify 
considerably: 
\bea
\LL_1 = {d^2\over d+1}\,\LL, &\quad & 
\LL_2 = 2d\,\LL,\\
\MM_1 = {d\over d+1}\,\LL, &\quad & 
\MM_2 = {2(d+1)(d+4)\over d+3}\,\LL.
\eea

If we take the long-time limit, i.e.,~$t\gg\tcorr$, 
the coefficients in~\eqref{Z_eq_closed} do not depend on time~$t$.
We can now use the inverse Fourier transform of~\eqref{Z_eq_closed} 
taken in this limit 
and the isotropic operators listed above to assemble 
the Fokker--Planck equation for the~PDF of the magnetic field.
This equation contains the desired corrections that are of first order 
in the velocity correlation time~$\tcorr$ and represent 
the first available manifestation of the finite-correlation-time 
effects. We have
\bea
\label{FPEq_expanded}
\dt P =  {\kbar_2\over2}
\(\LL - {1\over2}\,\tcorr\kbar_2\bigl[K_1\bigl(\LL_1+\MM_1\bigr) + K_2\LL_2 
+ \tK_2\MM_2\bigr]\)P,
\eea 
where the overall dimensional factor is 
\bea
\label{deff_kbar2}
\kbar_2 = 2\int_0^\infty\diff\tau\,\kappa_2(\tau),
\eea
and the coefficients~\cite{fnote_coeffs},
\bea
\label{deff_K1}
K_1 &=& {4\over\tcorr\kbar_2^2}\lim_{t\to\infty}
\int_0^t\diff t_1\int_{t_1}^t\diff t_2\int_0^{t_1}\diff t_3\,
\kappa_2(t-t_1)\kappa_2(t_2-t_3),\\
\label{deff_K2}
K_2 &=& {4\over\tcorr\kbar_2^2}\lim_{t\to\infty}
\int_0^t\diff t_1\int_{t_1}^t\diff t_2\int_{t_1}^{t_2}\diff t_3\,
\kappa_2(t-t_1)\kappa_2(t_2-t_3),\\
\label{deff_tK2}
\tK_2 &=& {4\over\tcorr\kbar_2^2}\lim_{t\to\infty}
\int_0^t\diff t_1\int_{t_1}^t\diff t_2\int_{t_1}^{t_2}\diff t_3\,
\kappa_4(t-t_1)\kappa_0(t_2-t_3),
\eea
are constants that depend on the particular shapes of 
the time-correlation functions~$\kappa_0(\tau)$, $\kappa_2(\tau)$, 
and~$\kappa_4(\tau)$. Such sensitive dependence is a new feature 
and represents a loss of universality with respect to the specific 
time-correlation profiles (cf.~Ref.~\cite{Boldyrev_tcorr}). 
As we have pointed out in~\ssecref{tau_expansion},  
the universality with respect to the functional form of 
the velocity correlator in space is also lost (this effect 
is incorporated into the coefficient~$\tK_2$).  

Let us also list the much more compact form that the Fokker--Planck 
equation~\exref{FPEq_expanded} assumes in the case of an incompressible 
velocity field:
\bea
\label{FPEq_expanded_inc}
\dt P =  {\kbar_2\over2}
\[1 - \tcorr\kbar_2 d\({1\over2}\,K_1 + K_2 
+ {(d+1)(d+4)\over d(d+3)}\,\tK_2\)\] \LL P.
\eea
Here it is especially manifest that the true expansion 
parameter in the problem is~$\tcorr\kbar_2 d$. 
This is a general statement that holds regardless of the 
degree of compressibility, as can be readily verified 
by counting powers of~$d$ in the general expressions for 
the operators~$\LL$, $\LL_1$, $\LL_2$, $\MM_1$, and~$\MM_2$ 
[formulas~\exref{deff_LL}--\exref{deff_MM2}].

It is evident that the distribution resulting 
from~\eqref{FPEq_expanded} is lognormal, which is a well-known fact 
in the kinematic-dynamo and passive-advection theory. 
Since we are interested in 
the quantitative description of the fast-dynamo effect, we will 
now proceed to calculate the growth rates of the moments of 
the magnetic field.

\subsection{The Dynamo Growth Rates}
\label{growth_rates_expanded}

The evolution of all moments of~$B$ can be determined 
from~\eqref{FPEq_expanded}. 
The $n$th moment is calculated according to
\bea
\label{def_Bn}
\<B^n\> = {2\pi^{d/2}\over\Gamma(d/2)}\int_0^\infty{\rm d}B\,B^{n+d-1}P(t;B).
\eea
Upon multiplying both sides of~\eqref{FPEq_expanded} by~$B^{d+n-1}$ and 
integrating over~$B$, we find that~$\<B^n\>$ satisfies:
\bea
\nonumber
\dt\<B^n\> &=& \gamma(n)\<B^n\>\\
\label{Bn_eq} 
&=& {\kbar_2\over2}\Bigl\{\Gamma(n) - 
\tcorr\kbar_2 d\bigl[K_1\Gamma_1(n) + K_2\Gamma_2(n) 
+ \tK_2\tGamma_2(n)\bigr]\Bigr\}\<B^n\>,
\eea
where the nondimensionalized zeroth-order growth rates 
are (cf.~Ref.~\cite{BS_metric}) 
\bea
\label{Gamma0}
\Gamma(n) = {d-1\over d+1}\,n\bigl[n+d + (n-1)\igamma\bigr],
\eea 
and the universal parts of the 
(negative) first-order corrections arising from the second- and 
fourth-order terms in the velocity correlator~\exref{xi_024} are
\bea
\label{Gamma1}
\Gamma_1(n) &=& {1\over2}\,{d-1\over d+1}\,n
\[\(n+d\)\(1+{2\igamma\over d^2}\) + \(n-1\)\,{\igamma^2\over d^2}\],\\
\label{Gamma2}
\Gamma_2(n) &=& {d-1\over d+1}\,n\(n+d\)\(1+{\igamma\over d^2}\),\\
\label{tGamma2}
\tGamma_2(n) &=& {(d-1)(d+4)\over d(d+3)}\,n
\[n+d + {n\over2}\({d^2+4d+2\over d(d+4)}-1\)\zeta\].
\eea 
We observe that, for $n=0$, 
$\Gamma=\Gamma_1=\Gamma_2=\tGamma_2=0$. This simply means that 
both zeroth- and first-order terms in the $\tau$~expansion 
preserve the normalization of the~PDF, i.e., our expansion 
is {\em conservative,} as it should be. 

In the incompressible flow, the total growth rate 
of the $n$th moment can be written in a more compact form:
\bea
\label{gamma_iso}
\gamma(n) = {\kbar_2\over2}\,{d-1\over d+1}\,n(n+d)
\l[1 - \tcorr\kbar_2 d\({1\over2}\,K_1 + K_2 
+ {(d+1)(d+4)\over d(d+3)}\,\tK_2\)\].
\eea

We see that the corrections to the growth rates 
of the magnetic-field moments are negative, so the growth 
rates are reduced. The amount of reduction depends on 
a variety of factors including the dimension of space, 
the order of the moment, the degree of compressibility,  
the functional form of 
the velocity correlator in time and space, and, of course, 
the velocity correlation time.  
Let us note that our general results derived for an arbitrarily 
compressible velocity field reveal no qualitatively essential 
effect of compressibility on the behavior of the first-order 
finite-correlation-time corrections to the dynamo growth rates 
in the diffusion-free regime. 
Compressibility of the flow simply leads to additive (and positive) 
corrections to the incompressible values of~$\Gamma(n)$, 
$\Gamma_1(n)$, $\Gamma_2(n)$, and~$\tGamma_2(n)$. Quantitatively, 
these corrections may affect the exact conditions for the 
break-down of the first-order approximation. For more discussion 
of the compressibility effects in the kinematic dynamo 
(with a $\delta$-correlated velocity field), we address 
the reader to Refs.~\cite{Rogachevskii_Kleeorin,BS_metric,SCMM_folding}. 

We remind the reader that here we have studied magnetic fluctuations 
{\em in the diffusion-free regime} and therefore dropped the term 
in the induction equation that is responsible for the resistive 
regularization. Such an approach is justified for plasmas with 
very large magnetic Prandtl numbers (e.g., the ISM or the prototogalaxy) 
and applies to the initial stage of the small-scale dynamo 
that lasts for a time of order~$t\sim\log\Pr$ that elapses 
before the magnetic fluctuations reach resistive 
scales~\cite{Kulsrud_lecture,SBK_review}. After that, or 
if the Prandtl number is of order unity or small (as is, e.g., 
the case for the Sun), resistive effects must be taken into account. 
In this case, the calculation of the moments of the 
magnetic field via the Fokker--Planck equation for its~PDF 
as presented in this Section does not apply because of the 
closure problem associated with the diffusion term 
[the equations for~$\tZ(t,\vx;\mu)$ and~$Z(t;\mu)$ do not close]. 
However, the general $\tau$-expansion method proposed in this paper 
can, in principle, be applied to multipoint correlators 
of the magnetic field, for which treating the diffusive case 
presents no conceptual difficulty. One-point moments can then 
be obtained by fusing the points at which the multipoint 
correlators are taken~(cf.~Refs.~\cite{Frisch_etal,fnote_diff_result}). 
Although it is the diffusive case that is studied in most 
numerical simulations, where $\Pr$ rarely exceeds~$100$, 
it is not necessarily the most relevant 
one in the context of the (proto)galactic dynamo, 
for which $\Pr\sim10^{14}\div 10^{22}$. Indeed, as we already 
pointed out in the Introduction, the initial (proto)galactic 
seed field may well be strong enough for the kinematic 
approximation to break down while the dynamo is still in 
the diffusion-free stage~\cite{Kulsrud_lecture}. 
If this is the case, the effect of magnetic diffusion 
must be studied in conjunction with nonlinear saturation 
of the magnetic fluctuations~\cite{Kinney_etal_2D}.

\section{A Physical Example: The One-Eddy Model} 
\label{sec_one_eddy}


In real astrophysical environments, such as the interstellar medium 
and the protogalactic plasmas, the magnetic fields are 
acted upon by a Kolmogorov-like turbulence with a fully developed 
inertial range about three decades wide ($\Re\sim 10^4$). 
While the velocities of the turbulent eddies excited by the 
Kolmogorov cascade decrease with the scale of the eddy, 
the velocity gradients increase (see, e.g.,~Ref.~\cite{Frisch_book}). 
Therefore, the dominant role in the process of amplification of
the small-scale magnetic fluctuations is played by the smallest 
eddies. With this circumstance 
in mind, one often considers, for modeling purposes, a synthetic 
incompressible turbulent velocity field consisting of eddies 
all of which have the same fixed size but random isotropic 
orientation (for detailed discussions of the galactic and protogalactic 
dynamo, we refer the reader 
to~Refs.~\cite{KA,Kulsrud_etal_proto,Kulsrud_lecture,SBK_review}).  
In this Section, we will present a brief discussion of the implications 
of the $\tau$-expansion theory developed in~\secref{sec_KT_dynamo} 
for such a model problem, which will 
henceforth be referred to as the {\em one-eddy model}. 

The velocity field in the one-eddy model is specified as follows:
\bea
\xi^i(t,\vx) &=& \intk e^{i\vk\cdot\vx}\xi^i(t,\vk), 
\eea
where the Fourier modes~$\xi^i(t,\vk)$ are random variables 
that satisfy  
\bea
\bigl<\xi^i(t,\vk)\xi^j(t',\vk')\bigr> &=& (2\pi)^d\delta(\vk+\vk') 
\(\delta^{ij} - {k_i k_j\over k^2}\)\delta(k-k_0)\kappa(t-t'). 
\eea
In this case, $\kappa_2(\tau)\propto\kappa(\tau)$, and, upon using 
the relations listed in~\Apref{ap_k_space}, we~get 
\bea
\label{kappa0_kappa2}
\kappa_0(\tau) &=& {1\over k_0^2}\,{(d-1)(d+2)\over d+1}\,\kappa_2(\tau),\\
\label{kappa4_kappa2}
\kappa_4(\tau) &=& k_0^2\,{d+3\over2(d+4)(d+1)}\,\kappa_2(\tau). 
\eea
Let us specify a plausible velocity time-correlation profile:
\bea
\label{kappa2_OU}
\kappa_2(\tau) = {\kbar_2\over2\tcorr}\,
\exp\(-{|\tau|\over\tcorr}\).
\eea
For this correlation function, which corresponds, for example, 
to the well-known Ornstein--Uhlenbeck random process 
(see, e.g.,~Ref.~\cite{Almighty_Chance}), 
the coefficients of the $\tau$~expansion~\exref{gamma_iso} 
are~$K_1=K_2=1/2$. 
The relations~\exref{kappa0_kappa2} and~\exref{kappa4_kappa2} 
provide the value of~$\tK_2$: 
\bea
\label{tK2_k}
\tK_2 = {(d-1)(d+2)(d+3)\over2(d+1)^2(d+4)}\,K_2. 
\eea

Let us define the ``eddy-turnover'' time~$\teddy\sim(k_0\xi)^{-1}$ 
of such a velocity field according to the following relation: 
\bea
\label{deff_teddy_k} 
{1\over\teddy^2} = k_0^2\,{1\over\tcorr} 
\int_{-\infty}^{+\infty}\diff\tau\,\kappa^{ii}(\tau,\vy=0) 
= k_0^2\,d\,{\kbar_0\over\tcorr} 
= {d(d-1)(d+2)\over d+1}\,{\kbar_2\over\tcorr},  
\eea 
where we have used~\eqref{kappa0_kappa2} to express 
$\kbar_0$ in terms of~$\kbar_2$. 
Note that the same expression is obtained if 
$\teddy\sim(\nabla\vxi:\nabla\vxi)^{-1/2}$~is 
formally defined in terms of the velocity gradients 
(without recourse to the one-eddy model):
\bea
\label{deff_teddy_x}
{1\over\teddy^2} = {1\over\tcorr}
\int_{-\infty}^{+\infty}\diff\tau\,|\kappa^{ii}_{,jj}(\tau,\vy=0)|
= {d(d-1)(d+2)\over d+1}\,{\kbar_2\over\tcorr}.
\eea
We recall that the zeroth-order growth rate~$\gamma_0$ of the 
magnetic-fluctuation energy~$\<B^2\>$ is~[see formula~\exref{gamma_iso}] 
\bea
\label{deff_gamma} 
\gamma_0 = {(d-1)(d+2)\over d+1}\,\kbar_2.
\eea  
Formulas~\exref{deff_teddy_k} and~\exref{deff_gamma} then imply 
\bea
\label{small_param}
\({\tcorr\over\teddy}\)^2 = \tcorr\gamma_0 d 
= {(d-1)(d+2)\over d+1}\,\tcorr\kbar_2 d.
\eea
We have established a correspondence between the small 
parameter that has arisen in our expansion of the dynamo 
growth rates and the ``physical'' small parameter, which is
the ratio of the correlation and eddy-turnover times. 
Of course, the above expression hinges on 
the definitions~\exref{deff_teddy_k} or~\exref{deff_teddy_x} of~$\teddy$. 
A simple physical argument can be made in favor of these definitions 
and the resulting formula~\exref{small_param}. 
Namely, let us observe that when~$\tcorr\sim\teddy$ the eddy 
only stretches the magnetic field line in one of the~$d$ 
available directions during one turnover time, 
whence~$\tcorr\gamma \sim 1/d$. The same estimate follows 
from the formula~\exref{small_param}.

Let us now evaluate the first-order correction to the growth rate 
of the magnetic energy. In the one-eddy model, one gets, upon using 
formulas~\exref{gamma_iso} and~\exref{tK2_k} and taking~$K_1=K_2=1/2$ 
for the Ornstein--Uhlenbeck time-correlation profile~\exref{kappa2_OU}, 
\bea
\label{exp_gamma}
\gamma = \gamma(2) = \gamma_0\bigl(1 - C_d\tcorr\gamma_0 d\bigr), 
\qquad C_d = {2d(d+1)-1\over2d(d-1)(d+2)}. 
\eea
We note that in three dimensions, $C_d=23/60\simeq 40\%$. 
When~$\tcorr\sim\teddy$, we have $\tcorr\gamma_0 d\sim 1$, 
and the resulting growth-rate reduction of~$\sim40\%$ is in 
a good qualitative agreement with the available numerical 
results~\cite{Chandran_tcorr,Kinney_etal_2D,Chou}. Of course, 
as we have already stressed in the Introduction, our 
$\tau$~expansion is not designed for the case of~$\tcorr\sim\teddy$, 
so the fact that it gives a fairly reasonable prediction should not be 
considered as an adequate quantitative corroboration 
of our theory. At best, one might conclude that the 
first-order expansion is well behaved for not-too-small values 
of the expansion parameter. 

Let us emphasize, however, that 
such a well-behaved expression has resulted from a number 
of essentially arbitrary (albeit physically reasonable)  
specifications of the parameters involved in the $\tau$~expansion. 
One of the most physically important points that we have tried 
to make in this work is, in fact, that the inclusion of 
finite-correlation-time effects leads to {\em nonuniversal} 
statistics, so the quantitative predictions of the theory 
can and will change appreciably if such factors as 
the shapes of the time-correlation profiles are changed. 
Namely, one would obtain expressions of the form~\exref{exp_gamma} 
with different values of the coefficient~$C_d$. 
For sufficiently large values of~$\tcorr\gamma_0 d$, 
the validity of the expansion~\exref{exp_gamma} will break down, 
and the expression in the brackets may even become negative. 
However, the following heuristic argument can be envisioned 
in this context. 

Let us recall that the finite-correlation-time effect 
was due to the presence of time-history integrals such as 
those that appear in the equations~\exref{Z_eq}, 
\exref{G_eq}, and~\exref{G_eq_1}. 
The first-order corrections in the Fokker--Planck 
equation~\exref{FPEq_expanded} arose from systematically 
approximating the time evolution of the statistical quantities 
[response functions and characteristic function~$Z(t;\mu)$] 
that entered these time-history integrals. 
The corrected (``true'') value of~$\gamma$ represents, 
in a rough way, the rate at which these quantities change. 
It would appear then that a better estimate of~$\gamma$ 
would be obtained if~$\gamma_0$ in the first-order term in 
the brackets in~\eqref{exp_gamma} were replaced with 
the corrected value~$\gamma$. With this caveat, we would 
find~that~\cite{fnote_Pade}
\bea
\gamma = {\gamma_0\over 1 + C_d\tcorr\gamma_0 d}. 
\eea
To first order, this formula is equally accurate as~\eqref{exp_gamma}. 
However, it better represents the fact that, 
as $\tcorr\gamma_0 d$~increases, the corrected value of 
of the growth rate should be expected to saturate~\cite{Kinney_etal_2D}. 
Of course, such considerations cannot substitute for an adequate 
nonperturbative theory of the passive advection and kinematic dynamo 
in finite-time-correlated flows, which remains an open problem.

\section*{Acknowledgments}

The authors would like thank S.~A.~Boldyrev 
for extensive and very fruitful 
discussions of the physics and the formalism of 
the finite-time-correlated kinematic dynamo problem. 
Both the substance and the style of the presentation 
have benefited from suggestions made by J.~A.~Krommes who 
read an earlier manuscript of this work. 
We are also grateful to G.~Falkovich, V.~Lebedev, S.~Cowley, 
and the anonymous referee for several useful comments. 

This work was supported by the 
U.~S.~Department of Energy under Contract~No.~DE-AC02-76-CHO-3073.

\appendix

\section{Small-Scale-Expansion Coefficients of the Velocity Correlation 
Tensor in Terms of Velocity Spectra} 
\label{ap_k_space}

In this Appendix, we list the basic formulas that relate the coefficients 
of the small-scale expansion~\exref{xi_024} of the velocity correlation 
tensor to the spectral characteristics of the velocity field. 
These relations allow one to apply the results on the small-$\tcorr$ 
expansion obtained in~\secref{sec_KT_dynamo} to velocity fields 
that are specified in the Fourier, rather than configuration, space. 
They also provide a set of consistency constraints that must be respected 
when the specific functional forms of~$\kappa_0(\tau)$, $\kappa_2(\tau)$, 
and~$\kappa_4(\tau)$ are chosen. 

Let the advecting velocity field be given as a sum of spatial 
Fourier modes, 
\bea
\label{FT_inverse}
\xi^i(t,\vx) &=& \intk e^{i\vk\cdot\vx}\xi^i(t,\vk).
\eea
and let the Fourier coefficients~$\xi^i(t,\vk)$ be random variables 
that satisfy  
\bea
\label{kappa_k}
\bigl<\xi^i(t,\vk)\xi^j(t',\vk')\bigr> &=& (2\pi)^d\delta(\vk+\vk') 
\[\kappa(k,t-t')\delta^{ij} + \tkappa(k,t-t')\,{k_i k_j\over k^2}\].
\eea
For the incompressible flows,~$\tkappa(k,\tau)=-\kappa(k,\tau)$; 
for the irrotational ones,~$\kappa(k,\tau)=0$. 
The coefficients of the expansion~\exref{xi_024} can then be expressed 
as follows:
\bea
\label{kappa0_k}
\kappa_0(\tau) &=& {1\over d}\intk \bigl[d\kappa(k,\tau) + 
\tkappa(k,\tau)\bigr],\\
\kappa_2(\tau) &=& {1\over d(d+2)}\intk k^{2}\bigl[(d+2)\kappa(k,\tau) + 
\tkappa(k,\tau)\bigr],\\
\kappa_4(\tau) &=& {1\over 2d(d+2)(d+4)}\intk k^4\bigl[(d+4)\kappa(k,\tau) + 
\tkappa(k,\tau)\bigr],\\
a(\tau) &=& \kappa_2(\tau)^{-1}{1\over d(d+2)}\intk k^{2}\tkappa(k,\tau),\\ 
\label{b_k}
b(\tau) &=& \kappa_4(\tau)^{-1}{1\over d(d+2)(d+4)}\intk k^{4}\tkappa(k,\tau),
\eea 
where the $\vk$-space integrations of radial functions can, of course, 
be written more explicitly~as 
\bea
\intk = \Idk k^{d-1}, \qquad S_d={2\pi^{d/2}\over \Gamma(d/2)}.
\eea
The derivation of the above relations is straightforward and 
based on the expressions for the correlation functions 
of isotropic fields in configuration space in terms of their spectra. 
For the 3-D~case, these expressions can be found in~Ref.~\cite{Monin_Yaglom}. 
A detailed derivation of the formulas~\exref{kappa0_k}-\exref{b_k} 
for the $d$-dimensional case is also given in 
Appendix~A of~Ref.~\cite{SBK_review}.

\section{Second-Order Response Functions}
\label{ap_response2}

In this Appendix, we provide the zeroth-order expressions 
for the second-order response functions that we used 
in~\ssecref{tau_expansion}. They are all derived in the same 
fashion: \eqref{Ztilde_eq}~is formally integrated, functional 
derivatives of it are taken with respect to the velocity 
field~$\xi^i$ or its gradients~$\xi^i_{,k}$ at the appropriate 
moments, the result is averaged, and the causality property 
of the response functions is used. Here we simply list 
the results. 

When~$t_2\ge t_1$, we have, {\em to zeroth order in the correlation 
time~$\tcorr$,}
\bea
\nonumber
G^{\a_1}_{\b_1\b_2}(t',\vx|t_1,\vx_1;t_2,\vx_2) = 
G^{\a_1}_{\b_1\b_2}(t_2,\vx|t_1,\vx_1;t_2,\vx_2)
\qquad\qquad\\
= -\delta(\vx-\vx_2) G^{\a_1}_{\b_1,\b_2}(t_2,\vx|t_1,\vx_1) 
+ \[{\d\over\d x^n}\,\delta(\vx-\vx_2)\]\L^n_{\b_2}
G^{\a_1}_{\b_1}(t_2,\vx|t_1,\vx_1),
\label{Ga1b1b2}\\
\nonumber
G^{\a_1\a_2}_{\b_1\b_2}(t',\vx|t_1,\vx_1;t_2,\vx_2) = 
G^{\a_1\a_2}_{\b_1\b_2}(t_2,\vx|t_1,\vx_1;t_2,\vx_2) 
\qquad\qquad\\
= -\Delta^{\a_2}(\vx-\vx_2) G^{\a_1}_{\b_1,\b_2}(t_2,\vx|t_1,\vx_1) 
+ \delta(\vx-\vx_2)\L^{\a_2}_{\b_2}
G^{\a_1}_{\b_1}(t_2,\vx|t_1,\vx_1),
\label{Ga1a2b1b2}
\eea 
where we have introduced the following notation: by definition, 
\bea
{\delta\xi^m(t,\vx)\over\delta\xi^{\b_2}_{,\a_2}(t_2,\vx_2)} = 
\delta^m_{\b_2}\delta(t-t_2)\Delta^{\a_2}(\vx-\vx_2).
\eea
The function~$\Delta^{\a_2}(\vx-\vx_2)$ is nonrandom and has the following 
property, which will be all that we need to know 
about~it:
\bea
{\d\over\d x^n}\,\Delta^{\a_2}(\vx-\vx_2) = 
\delta^{\a_2}_n\delta(\vx-\vx_2).
\eea

When~$t_1>t_2$, the expressions~\exref{Ga1b1b2} and~\exref{Ga1a2b1b2}
vanish by causality, so we have to flip the order of functional 
differentiation:
\bea
\nonumber
G^{\a_1}_{\b_1\b_2}(t',\vx|t_1,\vx_1;t_2,\vx_2) = 
G^{\ \ \a_1}_{\b_2\b_1}(t_1,\vx|t_2,\vx_2;t_1,\vx_1) 
\qquad\qquad\\
= -\Delta^{\a_1}(\vx-\vx_1) G_{\b_2,\b_1}(t_1,\vx|t_2,\vx_2) 
+ \delta(\vx-\vx_1)\L^{\a_1}_{\b_1}
G_{\b_2}(t_1,\vx|t_2,\vx_2),
\label{Ga1b2b1}\\
\nonumber
G^{\a_1\a_2}_{\b_1\b_2}(t',\vx|t_1,\vx_1;t_2,\vx_2) = 
G^{\a_2\a_1}_{\b_2\b_1}(t_1,\vx|t_2,\vx_2;t_1,\vx_1) 
\qquad\qquad\\
= -\Delta^{\a_1}(\vx-\vx_1) G^{\a_2}_{\b_2,\b_1}(t_1,\vx|t_2,\vx_2) 
+ \delta(\vx-\vx_1)\L^{\a_1}_{\b_1}
G^{\a_2}_{\b_2}(t_1,\vx|t_2,\vx_2).
\label{Ga2a1b2b1}
\eea
In~\eqref{Ga1b2b1}, the following obvious notation was used:
\bea
G^{\ \ \a_1}_{\b_2\b_1}(t_1,\vx|t_2,\vx_2;t_1,\vx_1) 
= \<\delta^2\tZ(t_1,\vx)\over
\delta\xi^{\b_2}(t_2,\vx_2)\delta\xi^{\b_1}_{,\a_1}(t_1,\vx_1)\>,
\eea
and a new first-order response function appeared:
\bea
G_{\b_2}(t_1,\vx|t_2,\vx_2) 
= \<{\delta\tZ(t_1,\vx)\over\delta\xi^{\b_2}(t_2,\vx_2)}\>.
\eea
The equal-time form of this function is
\bea
\label{Gb2_same}
G_{\b_2}(t_2,\vx|t_2,\vx_2) 
= \[{\d\over\d x^n}\,\delta(\vx-\vx_2)\]\L^n_{\b_2} Z(t_2). 
\eea

The first-order response functions that appear in 
the formulas~\exref{Ga1b1b2}, \exref{Ga1a2b1b2}, \exref{Ga1b2b1}, 
and~\exref{Ga2a1b2b1} can be written as their equal-time 
values~\exref{G_same} and~\exref{Gb2_same} plus first-order terms. 
To zeroth order, we have therefore: for~$t_2\ge t_1$,
\bea
\nonumber
G^{\a_1}_{\b_1\b_2}(t',\vx|t_1,\vx_1;t_2,\vx_2) &=& 
-\[{\d\over\d x^{\b_2}}\,\delta(\vx-\vx_1)\]\delta(\vx-\vx_2) 
\L^{\a_1}_{\b_1} Z(t_1)\\
&+& \delta(\vx-\vx_1)\[{\d\over\d x^n}\,\delta(\vx-\vx_2)\]
\L^n_{\b_2}\L^{\a_1}_{\b_1} Z(t_1),\\
\nonumber
G^{\a_1\a_2}_{\b_1\b_2}(t',\vx|t_1,\vx_1;t_2,\vx_2) &=& 
-\[{\d\over\d x^{\b_2}}\,\delta(\vx-\vx_1)\]\Delta^{\a_2}(\vx-\vx_2)
\L^{\a_1}_{\b_1} Z(t_1)\\ 
&+& \delta(\vx-\vx_1)\delta(\vx-\vx_2)
\L^{\a_2}_{\b_2}\L^{\a_1}_{\b_1} Z(t_1);
\eea
for~$t_1\ge t_2$,
\bea
\nonumber
G^{\a_1}_{\b_1\b_2}(t',\vx|t_1,\vx_1;t_2,\vx_2) &=& 
-\Delta^{\a_1}(\vx-\vx_1)\[{\d^2\over\d x^{\b_1}\d x^n}\,\delta(\vx-\vx_2)\] 
\L^n_{\b_2} Z(t_2)\\
&+& \delta(\vx-\vx_1)\[{\d\over\d x^n}\,\delta(\vx-\vx_2)\]
\L^{\a_1}_{\b_1}\L^n_{\b_2} Z(t_2),\\
\nonumber
G^{\a_1\a_2}_{\b_1\b_2}(t',\vx|t_1,\vx_1;t_2,\vx_2) &=& 
-\Delta^{\a_1}(\vx-\vx_1)\[{\d\over\d x^{\b_1}}\,\delta(\vx-\vx_2)\] 
\L^{\a_2}_{\b_2} Z(t_2)\\
&+& \delta(\vx-\vx_1)\delta(\vx-\vx_2)
\L^{\a_1}_{\b_1}\L^{\a_2}_{\b_2} Z(t_2).
\eea

These expressions must be substituted into~\eqref{G_eq}.
The volume integrals with respect to~$\vx_2$ can be done, 
taking into account the extremely useful fact that all 
odd spatial derivatives of the velocity 
correlator~$\kappa^{ij}(\tau,\vy)$ vanish at 
the origin (at~$\vy=0$).
The results are: for~$t_2\ge t_1$,
\bea
\nonumber
\intx_2\,\kappa^{m\b_2}(t'-t_2,\vx-\vx_2)
G^{\a_1}_{\b_1\b_2,m}(t',\vx|t_1,\vx_1;t_2,\vx_2)\qquad\qquad\\ 
=\,\[-{\d^2\delta(\vx-\vx_1)\over\d x^{\b_2}\d x^m}\,
\kappa^{m\b_2}(t'-t_2,0) 
+ \delta(\vx-\vx_1)\,
\kappa^{m\b_2}_{,mn}(t'-t_2,0)\,
\L^n_{\b_2}\]\L^{\a_1}_{\b_1} Z(t_1),\\
\nonumber
\intx_2\,\kappa^{m\b_2}_{,n\a_2}(t'-t_2,\vx-\vx_2)
G^{\a_1\a_2}_{\b_1\b_2}(t',\vx|t_1,\vx_1;t_2,\vx_2)\qquad\qquad\\
=\,\delta(\vx-\vx_1)\,
\kappa^{m\b_2}_{,n\a_2}(t'-t_2,0)\,\L^n_m
\L^{\a_2}_{\b_2}\L^{\a_1}_{\b_1} Z(t_1);\qquad\qquad\qquad
\eea
for~$t_1\ge t_2$,
\bea
\nonumber
\intx_2\,\kappa^{m\b_2}(t'-t_2,\vx-\vx_2)
G^{\a_1}_{\b_1\b_2,m}(t',\vx|t_1,\vx_1;t_2,\vx_2)\qquad\qquad\\
=\,\delta(\vx-\vx_1)
\[-\kappa^{\a_1\b_2}_{,\b_1n}(t'-t_2,0) 
+ \kappa^{m\b_2}_{,mn}(t'-t_2,0)\,
\L^{\a_1}_{\b_1}\]\L^n_{\b_2} Z(t_2),\qquad\\ 
\nonumber
\intx_2\,\kappa^{m\b_2}_{,n\a_2}(t'-t_2,\vx-\vx_2)
G^{\a_1\a_2}_{\b_1\b_2}(t',\vx|t_1,\vx_1;t_2,\vx_2)\qquad\qquad\\ 
=\,\delta(\vx-\vx_1)\,
\kappa^{m\b_2}_{,n\a_2}(t'-t_2,0)\,\L^n_m
\L^{\a_1}_{\b_1}\L^{\a_2}_{\b_2} Z(t_2).\qquad\qquad\qquad 
\eea
With the aid of these expressions and~\eqref{G_eq}, one 
obtains~\eqref{G_eq_1} of~\ssecref{tau_expansion}.


\begin{thebibliography}{99}
\bibitem{fnote_KT_fusing_pts}
See the remark on the Kliatskin--Tatraskii 
method at the end of~\ssecref{KT_method_exp}. The reason for 
the discrepancy between their method and ours is rooted precisely 
in such illegitimate fusing of points.
\bibitem{fnote_operators} 
These operators are 
the generalized versions of the operators~$\LL_1$ and~$\LL_2$ 
of~\ssecref{KT_method_exp}.
\bibitem{fnote_coeffs}
The coefficients~$K_1$ and~$K_2$ 
are essentially the same as their namesakes that appeared 
in~\ssecref{KT_method_exp} (with~$\kbar$ replaced by~$\kbar_2$). 
The coefficient~$\tK_2$ would be proportional to~$K_2$ 
if we assumed that the time-correlation properties of the velocity 
correlator~$\kappa^{ij}(\tau,\vy)$ were embodied in a single 
scalar function, so that 
$\kappa_0(\tau)\propto\kappa_2(\tau)$ and 
$\kappa_4(\tau)\propto\kappa_2(\tau)$.
\bibitem{fnote_Pade}
Formally speaking, this formula 
represents the $(0,1)$-term of the Pad\'e series 
corresponding to the $\tau$~expansion. Whether the Pad\'e 
summation method applied to the $\tau$~expansion results in 
a good convergent approximation of the finite-correlation-time 
effect is an issue that will not be investigated here.
\end{thebibliography}
\end{document}